\documentclass[12pt]{iopart}

\usepackage{cite}
\usepackage{graphicx}
\usepackage{setstack}
\usepackage{iopams}

\begin{document}

\title{Nanospintronics with carbon nanotubes}
\author{Audrey Cottet$^{1}$, Takis Kontos$^{2}$, Sangeeta Sahoo$^{3}$, Hon Tin Man$^{4}$, Mahn-Soo Choi$^{5}$, Wolfgang Belzig$^{6}$,
Christoph Bruder$^{3}$, Alberto Morpurgo$^{4}$ and Christian
Sch\"onenberger$^{3}$}
\address{
{$^{1}$Laboratoire de Physique des Solides, B\^atiment 510,
Universit\'e
Paris Sud 91405, Orsay Cedex, France}\\
{$^{2}$ Laboratoire Pierre Aigrain, Ecole Normale Sup\'erieure, 24,
rue Lhomond, 75231 Paris Cedex 05, France}\\
{$^{3}$Institute of Physics, University of Basel, Klingelbergstr.
82, CH-4056 Basel, Switzerland}\\
{$^{4}$Kavli Institute of NanoScience Delft, Faculty of Applied
Sciences, Delft University of Technology, Lorentzweg 1, 2628 CJ Delft, The Netherlands}\\
{$^{5}$Department of Physics, Korea University, Seoul 136-701, Korea}\\
{$^{6}$Departement of Physics, University of Konstanz, M703, D-78457
Konstanz, Germany}}
\eads{\mailto{cottet@lps.u-psud.fr},\mailto{kontos@lpa.ens.fr}}
\begin{abstract}
One of the actual challenges of spintronics is the realization of a
spin-transistor allowing to control spin transport through an
electrostatic gate. In this review, we report on different
experiments which demonstrate a gate control of spin transport in a
carbon nanotube connected to ferromagnetic leads. We also discuss
some theoretical approaches which can be used to analyze spin
transport in these systems. We emphasize the roles of the
gate-tunable quasi-bound states inside the nanotube and the coherent
spin-dependent scattering at the interfaces between the nanotube and
its ferromagnetic contacts.
\end{abstract}
\date{\today}
 \maketitle

\section{Introduction : nanospintronics}

The quantum mechanical spin degree of freedom is now widely
exploited to control current transport in electronic devices. For
instance, the readout of magnetic hard disks is based on the
spin-valve effect, i.e. the tunability of a conductance through the
relative orientation of some ferromagnetic polarizations
\cite{Prinz}. However, realizing spin injection in nanostructures,
e.g. mesoscopic conductors or molecules, would allow to implement
further functionalities. For example, the realization of a
''spin-transistor'' would allow an electric field control of the
spin valve effect through an electrostatic gate
\cite{Datta,Schapers}. In this context, carbon nanotubes are
particularly interesting, because they should exhibit a long spin
life time  and can be contacted with ferromagnetic materials. In
this review, we present the state of the art regarding the
realization of spin-transistor-like devices with carbon nanotubes.
In section 2, we introduce the basics of the spin-valve effect. In
section 3, we present a theoretical description of spin-transport in
quantum wires with ferromagnetic contacts. We put a special emphasis
on the roles of the gate-tunable resonant states inside the wire and
the coherent spin-dependent scattering at the boundaries of the
wire. In section 4, we present the state of the art in contacting
carbon nanotubes with ferromagnetic materials, and evoke different
contact effects which could mimic spin-dependent transport
phenomena. In section 5, we review different experiments which have
demonstrated a gate control of spin-transport in carbon nanotubes so
far. Eventually, we give some conclusions and perspectives in
section 6.

\section{The spin-valve geometry}

\begin{figure}[pth]
\centering\includegraphics[height=0.95\linewidth
,angle=-90,clip]{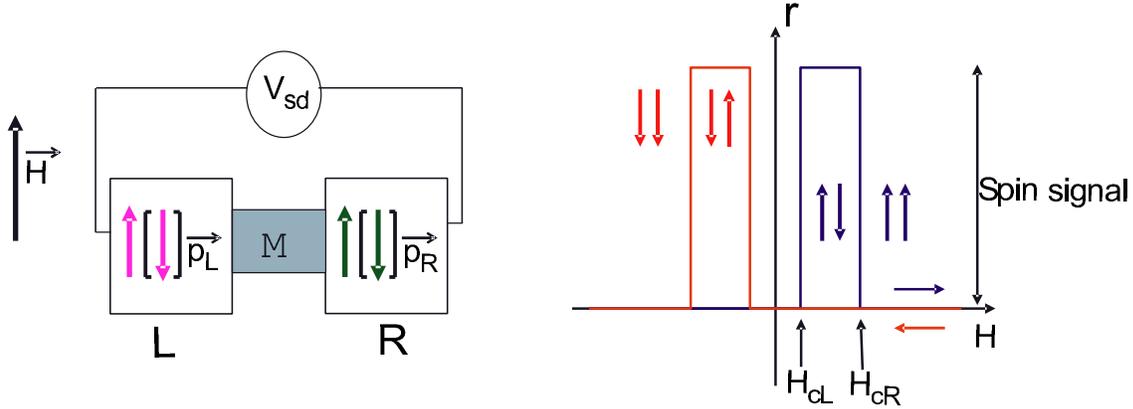}\caption{Left: Electrical diagram of a
circuit with the spin-valve geometry. The element M is connected to
two ferromagnetic leads $L$ and $R$, in which the electronic density
of states has magnetic polarizations $\vec{p}_{L}$ and
$\vec{p}_{R}$. The three elements in series form a spin-valve, which
is voltage-biased with a source-drain voltage $V_{sd}$. A magnetic
field $\overrightarrow{H}$ is applied to the circuit. Right: Typical
shape of the resistance curve $r(H)$ measured in the spin-valve
while increasing (blue line) and then decreasing (red line) $H$.
Since the two contacts $L$ and $R$ have different coercive fields
$H_{cL}$ and $H_{cR}$, it is possible to reverse selectively the
directions of $\vec{p}_{L}$ and $\vec{p}_{R}$ during this cycle.
This introduces an hysteretic pattern in the
$r(H)$ curve, from which the value of the magnetoresistance $MR=(G^{P}%
-G^{AP})/G^{AP}$ of the spin-valve can be obtained. Here, one has $MR>0$. }%
\label{fig:SpinValve}%
\end{figure}

The most standard method to inject or detect spins in an insulating
or conducting element $M$ is to use the spin-valve geometry
\cite{Baibich:88,Binasch:89}, in which $M$ is connected to two
ferromagnetic leads $L$ and $R$ (figure \ref{fig:SpinValve}, left).
One has to measure the conductances $G^{P}$ and $G^{AP}$ of the spin
valve for lead magnetizations in the parallel ($P$) and antiparallel
($AP$) configurations. This requires to use two ferromagnets with
different coercive fields ($H_{cL}$ and $H_{cR}$ respectively) for
switching one magnetization with respect to the other with the help
of an external magnetic field $H$ (figure \ref{fig:SpinValve},
right). The spin signal or magnetoresistance is then defined as the
relative difference $MR=(G^{P}-G^{AP})/G^{AP}$.

Let us consider the situation in which the element placed between
the two ferromagnetic contacts is a tunneling barrier with a
transmission probability independent of energy \cite{Julliere}. This
case, usually referred to as Julli\`ere's model, describes the
principle of magnetic memories and magnetic read heads. From Fermi's
Golden rule, the transmission probability of the barrier for spins
$\sigma\in\{\uparrow,\downarrow\}$ is proportional to the electronic
densities of states at the Fermi energy
$N_{l,\sigma}=N_{l}(1+\sigma\eta_{l}p_{l})$ for spins $\sigma$ at
both contacts, with $l\in\{L,R\}$ and $\eta_{l}\in\{+1,-1\}$ the
direction of the magnetization at contact $l$. Here, $N_{l}$ is the
spin averaged density of states and $p_{l}$ the spin polarization in
contact $l$. The conductance $G^{P}$ of the barrier in the parallel
configuration is proportional to
$N_{L}N_{R}[(1+p_{L})(1+p_{R})+(1-p_{L})(1-p_{R})]$ whereas the
conductance
$G^{AP}$ in the anti-parallel configuration is proportional to $N_{L}%
N_{R}[(1+p_{L})(1-p_{R})+(1-p_{L})(1+p_{R})]$. This leads to
\[
MR=\frac{2p_{L}p_{R}}{1-p_{L}p_{R}}.%
\]
If the spin polarizations $p_{L}$ and $p_{R}$ have the same sign,
the magnetoresistance of the device is positive because the current
flowing in the antiparallel configuration is lower due to the
imbalance between $N_{L,\sigma}$ and $N_{R,\sigma}$.

In the following, we consider the case in which element $M$ is a
carbon nanotube. In contrast with Julli\`ere's model of a tunneling
barrier, the transmission of the nanotube cannot be considered as
constant with energy due to the existence of quasi-bound states
between the two ferromagnetic contacts. Furthermore, it is possible
to tune the energy of these bound states with an electrostatic gate.
This significantly modifies the behavior of carbon-nanotube-based
spin valves, as we are going to explain theoretically in the next
section.

\section{Spin transport in finite size quantum wires\label{coherent}}

Carbon nanotubes can display a large variety of behavior, depending
on their intrinsic properties and on the characteristics of their
electrical contacts. Even in the case of a clean nanotube (i.e. with
no structural defects), electronic transport can occur in different
regimes, depending on the transparency of the contacts. For high
contact resistances $R>h/e^{2}$, a nanotube can behave as a quantum
dot, in which Coulomb blockade determines the transport properties
\cite{Devoret:98}, whereas for low contact resistances $R<h/e^{2}$,
transport is mainly determined by quantum interference
\cite{bockrath:01}. Here, we will mainly consider these two
situations. For simplicity, we will model the nanotube as a one
dimensional quantum wire.

\subsection{Spin dependent transport in a non-interacting ballistic wire.
\label{ballistic}}

Although electron-electron interactions should be of primary
importance in one dimensional quantum wires, it is instructive to
consider first a non-interacting picture. In addition, as we will
see in section \ref{bip}, such a simplified picture captures the
main features of some available experiments.
\subsubsection{Transmission of a F-wire-F ballistic system}

\begin{figure}[pth]
\centering\includegraphics[height=0.75\linewidth,angle=-90]{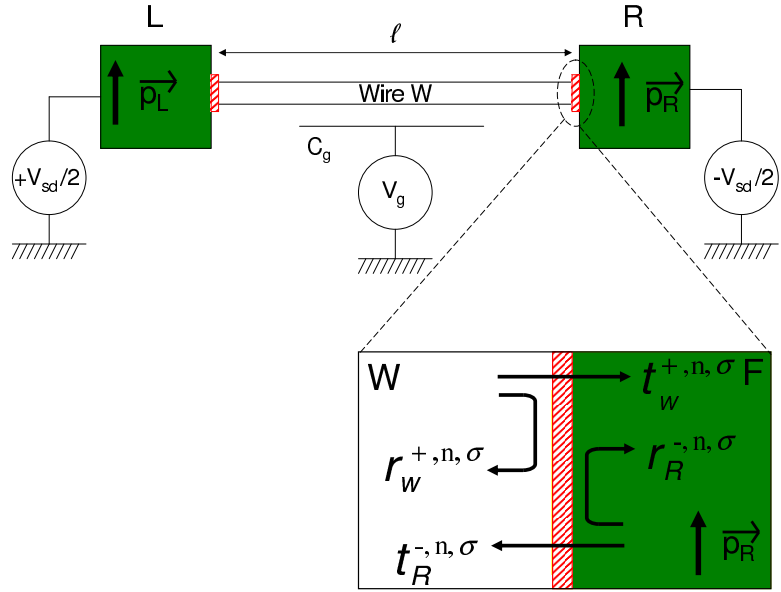}%
\caption{Electrical diagram of a ballistic wire $w$ with length
$\ell$ connected to ferromagnetic leads $L$ and $R$ with magnetic
polarizations $\vec{p}_{L}$ and $\vec{p}_{R}$. The wire is
voltage-biased with a source-drain voltage $V_{sd}$ and capacitively
coupled to a gate voltage source $V_{g}$. Inset: Scattering
description of the interface between the wire and a contact. We use
transmission and reflection amplitudes $t_{l}^{\epsilon,n,\sigma}$
and $r_{l}^{\epsilon,n,\sigma}$ for electrons with spin $\sigma$ of
transverse channel $n$, incident from element $l\in\{L,R,w\}$ with
direction $\epsilon$ ($\epsilon=+$ for right-going incident
electrons and $\epsilon=-$ for left-going
incident electrons)}%
\label{diagramm}%
\end{figure}

We consider the circuit of figure \ref{diagramm}, with W a
non-interacting ballistic wire with length $\ell$ contacted to two
ferromagnetic leads $L$ and $R$. In the non-interacting limit,
electronic transport through this device can be described using a
scattering approach \cite{Blanter}. As represented in the bottom
inset of figure \ref{diagramm}, this description involves complex
amplitudes of transmission and reflection
$t_{l}^{\epsilon,n,\sigma}$ and $r_{l}^{\epsilon,n,\sigma}$ for
electrons with spin $\sigma$ of transverse channel $n$, incident
from element $l\in\{L,R,w\}$ with direction $\epsilon$ ($\epsilon=+$
for right-going incident electrons and $\epsilon=-$ for left-going
incident electrons). Assuming that the different channels $n$ are
not coupled by interfacial scattering, the behavior of the device
only depends on the
transmission probabilities $T_{L(R)}^{n,\sigma}=|t_{L(R)}^{+(-),n,\sigma}%
|^{2}$ and on the reflection phases $\varphi
_{L}^{n,\sigma}=\arg(r_{w}^{-,n,\sigma})$ and $\varphi_{R}^{n,\sigma}%
=\arg(r_{w}^{+,n,\sigma})$ at the side of the wire. Indeed, the
conductance of
the circuit in configuration $P[AP]$ can be calculated from the expression%

\begin{equation}
G^{P[AP]}=G_{Q}\sum\nolimits_{n,\sigma}\int\nolimits_{-\infty}^{+\infty
}\mathbb{T}_{P[AP]}^{n,\sigma}(E)(-\partial f(E)/\partial
E)\label{Gballist}%
\end{equation}
where $f(E)=[1+\exp(E/k_{b}T)]^{-1}$ is the Fermi-Dirac distribution
and where
\begin{equation}
\mathbb{T}_{P[AP]}^{n,\sigma}=\frac{T_{L}^{n,\sigma}T_{R}^{n,\sigma}}{\left|
1-[(1-T_{L}^{n,\sigma})(1-T_{R}^{n,\sigma})]^{1/2}e^{i(\varphi_{L}^{n,\sigma
}+\varphi_{R}^{n,\sigma}+2\delta_{0}+\sigma \gamma^{n}_{H})}\right|  ^{2}}\label{ResTOT}%
\end{equation}
is the probability that an electron of channel $n$ with spin
$\sigma$ coming from lead $L(R)$ is transmitted to lead $R(L)$.
Here, we have introduced the orbital phase $\delta_{0}=\ell
k_{Fw}^{n}(V_{g})$ acquired by an electron upon crossing the wire
once and $\gamma^{n}_{H}=g \mu_{B} H \ell /\hbar v^{n}_{Fw}$, with
$H$ the external magnetic field. We call $k_{Fw}^{n}(V_{g})$ the
gate-dependent wave vector of electrons of channel $n$ inside the
wire, $v_{Fw}^{n}(V_{g})$ the corresponding Fermi velocity, $g$ the
Land\'e factor and $\mu_B$ the Bohr magneton. The denominator of
equation (\ref{ResTOT}) accounts for the existence of resonant
states which are due to multiple reflections
between the two contacts. These resonances lead to peaks in the $G^{P[AP]}%
(V_{g})$ curves. In the case of ferromagnetic contacts, the
interfacial scattering properties depend on spin and on the
configuration $c\in\{P,AP\}$ of the ferromagnetic electrodes (we
omit the index $c$ in $T_{L(R)}^{n,\sigma}$ and
$\varphi_{L(R)}^{n,\sigma}$ for brevity). In the following, we
investigate the effects of a finite spin polarization of the
tunneling rates $P_{l}^{n}\neq0$, and of a \textit{Spin-Dependence}
of Interfacial Phase Shifts (SDIPS), i.e.
$\Delta\varphi_{l}^{n}\neq0$, with
\[
T_{l}^{n,\sigma}=T_{l}^{n}(1+\eta_{l}\sigma P_{l}^{n})
\]%
\[
\varphi_{l}^{n,\sigma}=\varphi_{l}^{n}+\eta_{l}\sigma\frac{\Delta\varphi
_{l}^{n}}{2}.%
\]
for $l\in\{L,R\}$. Here, $\eta_{l}\in\{+1,-1\}$ denotes the
direction of the magnetization at contact $l$.

The quantum wires which we have in mind are carbon nanotubes. Two
different types of carbon nanotubes can be fabricated: Single-Wall
Nanotubes (SWNTs) and Multi-Wall Nanotubes (MWNTs). A SWNT consists
of a single graphene sheet that is rolled up into a cylinder. A MWNT
consists of a set of coaxially stacked graphene cylinders. In the
case of a SWNT, it is possible to have only two channels involved in
current transport at low voltages (the energy levels of SWNTs often
display a twofold degeneracy related to the K-K' degeneracy of the
energy bands of graphene \cite{Sapmaz:05,KK}). Assuming two
identical channels with no coupling, the behavior of such a nanotube
can be understood from the study of a one-channel quantum wire,
which is presented in section \ref{1channel} (the conductance of the
nanotube will be twice that of the single-channel quantum wire and
the magnetoresistance will be identical). For MWNTs, more channels
are generally involved in the low voltage electronic transport. We
will thus present in section \ref{multichannel} the case of a
quantum wire with several channels.

\subsubsection{Single channel case\label{1channel}}

In this section, we omit the channel index $n$. We assume that the
gate voltage $V_{g}$ induces a shift of the wire electrostatic
potential which is small compared the Fermi energy of the wire, i.e.
$\alpha V_{g}\ll E_{Fw}$, where $\alpha=C_{g}/C_{\Sigma}$ is the
ratio between the gate capacitance and the
total capacitance of the wire. In this limit, one finds $\delta_{0}%
=\ell k_{Fw}+(e\alpha V_{g}-E_{Fw})(\pi N_{Fw}/2)$ where $N_{Fw}=2
\ell/\pi\hbar v_{Fw}$ is the density of states in the wire, and
$k_{Fw}$ and $v_{Fw}$ are the Fermi wavevector and velocity in the
wire. Therefore, the resonant peaks in the $G^{P[AP]}(V_{g})$ curve
correspond to the cancelation of resonant energies of the form
\begin{eqnarray}
E_{P[AP]}^{\sigma,j}=(2\pi j-\varphi_{L}^{\sigma}-\varphi_{R}^{\sigma}%
-\sigma \gamma_{H})(\hbar v_{Fw}/2\ell )-e\alpha
V_{g},\label{eq:levels}
\end{eqnarray}
with $j\in\mathbb{Z}$.

\paragraph{Magnetoresistance of a 1-channel wire with no SDIPS}

In this paragraph, we investigate the behavior of the wire for
$\Delta
\varphi_{l}^{n}=0$, and thus define resonant energies $E^{j}=E_{P[AP]}%
^{\uparrow,j}=E_{P[AP]}^{\downarrow,j}$ for $\gamma_{H}=0$. From
equation (\ref{eq:levels}), the resonant peaks in the conductance
curves are spaced by $\Delta E=E^{j+1}-E^{j}=hv_{Fw}/2\ell$ which is
usually called the intrinsic level spacing of the wire. Figure
\ref{SDIPS} shows with black dashed lines the conductance
$G^{P}(V_{g})$ and the magnetoresistance $MR(V_{g})$ of a
one-channel wire. For convenience, we have plot the physical
quantities as a function of $\delta_{0}$ instead of the gate voltage
$V_{g}$. The conductance shows resonances with a $\pi$-periodicity
in $\delta_{0}$, corresponding to the intrinsic level spacing
$\Delta E$. Strikingly, the magnetoresistance can become negative
for certain values of $V_{g}$ which correspond to a resonance in
$G^{P}$. This is in contrast with the early Julli\`ere's model
evoked in section 2. In order to understand this situation (see
figure 3), it is convenient to consider the limit of low
transmissions $T_{l}\ll1$ , in which one can expand
$\mathbb{T}_{P[AP]}^{\sigma}$ around $E=E^{j}$ (see \cite{Blanter})
to obtain a Breit-Wigner-like formula \cite{Tsymbal}
\begin{equation}
\mathbb{T}_{P[AP]}^{\sigma}=\frac{T_{L}^{\sigma}T_{R}^{\sigma}}
{(\pi N_{Fw}[E-E^{j}])^{2}+(T_{L}^{\sigma}+T_{R}^{\sigma})^{2}/4}\label{BW}%
\end{equation}
Off resonance, i.e. when
$(E-E^{j})^{2}\gg(T_{L}^{\sigma}+T_{R}^{\sigma})/\pi N_{Fw}$, the
transmission probability $\mathbb{T}_{P[AP]}^{\sigma}$ of the
contact for electrons with spin $\sigma$ is proportional to
$T_{L}^{\sigma}T_{R}^{\sigma}$. This leads to
\[
MR=\frac{2P_{L}P_{R}}{1-P_{L}P_{R}}%
\]
like in Julli\`ere's model. At resonance, i.e. when $E=E^{j}$, the
situation is different. We will consider for simplicity the very
asymmetric case $T_{L}^{\sigma }\ll T_{R}^{\sigma}$. In this case,
equation (\ref{BW}) gives
$\mathbb{T}_{P[AP]}^{\sigma}=4T_{L}^{\sigma}%
/T_{R}^{\sigma}$, which leads to
\[
MR=-\frac{2P_{L}P_{R}}{1+P_{L}P_{R}}%
\]
Thus, it appears clearly that the change of sign in the $MR$ signal
is a direct consequence of the existence of quasi-bound states in
the wire.

\begin{figure}[pth]
\centering\includegraphics[width=0.75\linewidth,angle=-90]{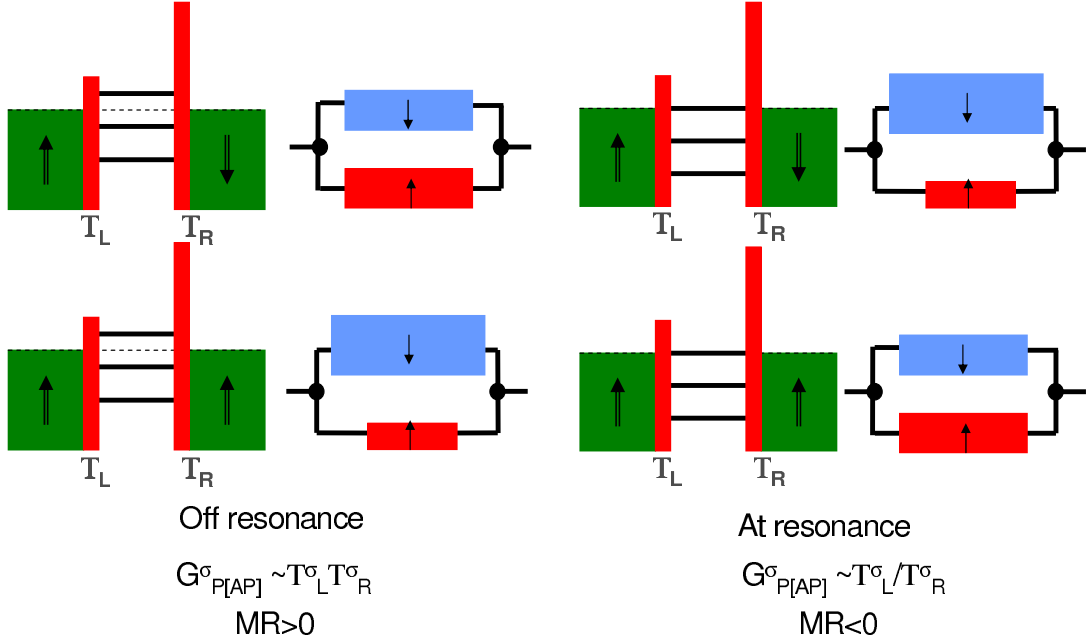}
\caption{Picture of the resonant tunneling mechanism for very
asymmetric barriers. Here, we assume $T_{L}^{\sigma}\ll
T_{R}^{\sigma}$ and $P_{L}=P_{R}$. A bigger resistance element
represents a higher resistance value. Off resonance (left panels),
the transmission probability $\mathbb{T}_{P[AP]}^{\sigma}$ of the
circuit for spins $\sigma$ scales with
$T_{L}^{\sigma}T_{R}^{\sigma}$ in the $P[AP]$ configuration. This
leads to a positive $MR$ like in Julli\`{e}re's model. On resonance
(right panels), $\mathbb{T}_{P[AP]}^{\sigma}$ scales with
$4T_{L}^{\sigma}/T_{R}^{\sigma}$, which leads to a negative $MR$.}
\label{Fig:resonant}
\end{figure}

\paragraph{Role of the Spin Dependence of Interfacial Phase Shifts (SDIPS).}

\begin{figure}[pth]
\centering\includegraphics[width=0.85\linewidth]{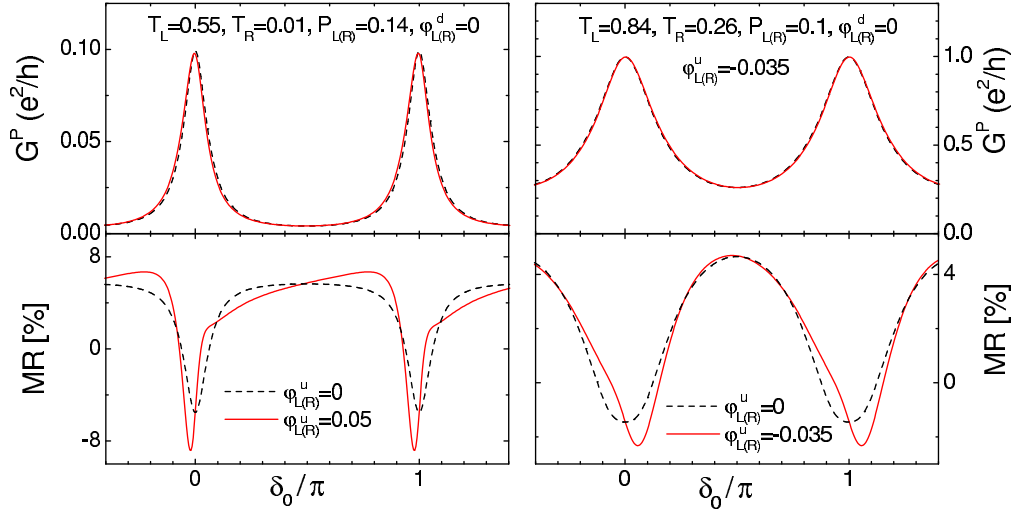}\caption{{Linear
conductance $G^{P}$ (top panels) and magnetoresistance $MR$ (bottom
panels) for a single channel wire, as a function of the
spin-averaged phase $\delta_{0}$ acquired by electrons upon crossing
the wire ($\delta_{0}$ is linear with $V_{g}$ in the limit
considered here, see text). We show the results in case of no SDIPS
(black dashed lines) and for a value of SDIPS finite but not
resolvable in the conductance curves (red full lines). The left and
right panels correspond to two different sets of parameters. When
the contacts have no SDIPS, the oscillations in $MR(\delta_{0})$ are
symmetric. A low SDIPS ($\left| \Delta\varphi^{P[AP]}\right|\lesssim
T_{L}+T_{R}$) can be detected qualitatively in the $MR(\delta_{0})$
curves because it breaks the symmetry of these oscillations. In
sections \ref{SWNTDelft} and \ref{SWNTBasel}, we will compare these
theoretical results with measurements done in SWNTs connected to
PdNi contacts. }}%
\label{SDIPS}%
\end{figure}

So far, we have assumed that the interfacial reflection phases of
electrons of channel $n$ coming from the wire were spin-independent,
i.e. $\varphi _{l}^{\uparrow}=\varphi_{l}^{\downarrow}$ for
$l\in\{L,R\}$. Nevertheless, the interface between a ferromagnet and
a non-magnetic material can scatter electrons with spin parallel or
antiparallel to the magnetization of the ferromagnet with different
phase shifts, because electrons are affected by a spin-dependent
scattering potential at this interface. This
\textit{Spin-Dependence} of Interfacial Phase Shifts (SDIPS) can
modify significantly the behavior of many different types of
mesoscopic circuits, like diffusive
ferromagnetic/normal/ferromagnetic spin valves \cite{FNF},
superconducting/ferromagnetic hybrid circuits \cite{supra}, Coulomb
blockade systems \cite{Wetzels, Cottet:06b} and Luttinger liquids
\cite{Balents}. Reference \cite{Cottet:06} has shown that
non-interacting ballistic wires can also be affected by the SDIPS.
From equation (\ref{eq:levels}), in case of a finite SDIPS, the
resonant energies of the 1-channel wire are spin-dependent. This
allows to define an SDIPS-induced effective field $h_{SDIPS}^{c}$
such that
\[
E_{c}^{\downarrow,j}-E_{c}^{\uparrow,j}=g\mu_{B}h_{SDIPS}^{c}=\frac
{\eta_{L}\Delta\varphi_{L}+\eta_{R}\Delta\varphi_{R}}{\pi N_{Fw}}%
\]
for $c\in\{P,AP\}$. Importantly, this effective field is
configuration-dependent. For instance, in the case of symmetric
barriers, one has $\Delta\varphi_{L}=\Delta\varphi_{R}$, thus
$h_{SDIPS}^{P}$ can be finite in the parallel case, but
$h_{SDIPS}^{AP}$ must vanish in the antiparallel case due to the
symmetry of the problem.

The effects of the effective field $h_{SDIPS}^{c}$ will depend on
its amplitude. Let us first imagine that the SDIPS effective field
is very strong in the parallel case, i.e. $\left|
\Delta\varphi_{L}+\Delta\varphi _{R}\right|  \gtrsim T_{L}+T_{R}$,
and that the barriers are symmetric, leading to $h_{SDIPS}^{AP}=0$
(one can imagine to obtain this situation by fabricating a symmetric
device with strongly spin-dependent barriers, by using e.g.
ferromagnetic insulators evaporated between the wire and the
metallic contacts). From equation (\ref{ResTOT}), this would allow
to resolve the level
spin-splitting $g\mu_{B}h_{SDIPS}^{P}$ in the conductance curve $G^{P}(V_{g}%
)$, and thus to obtain a shift of the conductance peaks from the $P$
to the $AP$ configurations. A giant $MR$ effect with a sign tunable
with $V_{g}$ would thus be obtained. This illustrates that a strong
SDIPS would be very useful for realizing an efficient control of
spin-dependent transport in nanowires. Nevertheless, since the
appropriate device fabrication has not been achieved yet, we refer
the reader to reference \cite{Cottet:06} for the detailed study of
the effects of a strong SDIPS and focus here on the case of a weaker
SDIPS. It is possible that a weak SDIPS occurs in actual
experiments. Indeed, when a standard ferromagnetic contact material
is evaporated directly on a wire, the interfacial scattering
potential which affects the electrons can already depend on spin.

Figure \ref{SDIPS}, red full lines, shows the conductance
$G^{P}(V_{g})$ (top panels) and the magnetoresistance $MR(V_{g})$
(bottom panels) for a device with a weak SDIPS. Although the
SDIPS-induced spin-splitting is too weak to be resolved in the
conductance curves for the parameters used here, it modifies
qualitatively the spin-valve behavior of the device. Indeed, when
there is no SDIPS, from equation (\ref{ResTOT}), the $MR(V_{g})$
oscillations are always symmetric with $V_{g}$. This symmetry is
broken by the SDIPS. This is due to the fact that,\ in the presence
of a weak SDIPS, the position of the global maximum corresponding to
$E_{c}^{\uparrow,j}$ and $E_{c}^{\downarrow,j}$ is different for
$c=P$ and $c=AP$. This effect provides a qualitative way to detect
the presence of a finite SDIPS in the circuit.

\subsubsection{Multichannel case.\label{multichannel}}

For MWNTs, it is usually assumed that transport occurs mainly
through the outer shell \cite{Bachtold:99}. However, since the
diameter of MWNTs is larger than that of SWNTs, the spacing between
the 1D subbands of the outer shell is lower \cite{Saito:98}. As a
consequence, a multichannel description is \textit{a priori} needed
is one wants to account for the $MR$. We have evaluated the
conductance and the $MR$ from equations (\ref{Gballist}) and
(\ref{ResTOT}), for a MWNT\ with two ferromagnetic contacts. In the
simple case where there is no subband mixing, one can determine the
transmission $\mathbb{T}_{P[AP]}^{n,\sigma}(E)$ occurring in these
equations via the wave vector
$k_{Fw}^{n}(V_{g})=k_{Fw}+\sqrt{(e\alpha V_{g}-E_{Fw})^{2}/(\hbar
v_{Fw})^{2}-n^{2}/R_{nt}^{2}}$, where $R_{nt}$ is the radius of the
MWNT \cite{Egger:05}. For a radius $R_{nt}=2.7nm$, the subband
spacing amounts to $\sim180meV$. As the Fermi energy shift of MWNTs
due to surface adsorbates can be as high as $\sim1eV$
\cite{Kruger:01}, up to 6 subbands can contribute to charge and spin
transport. We have thus taken into account 6 subbands in the
calculation.
\begin{figure}[pth]
\centering\includegraphics[width=0.65\linewidth]{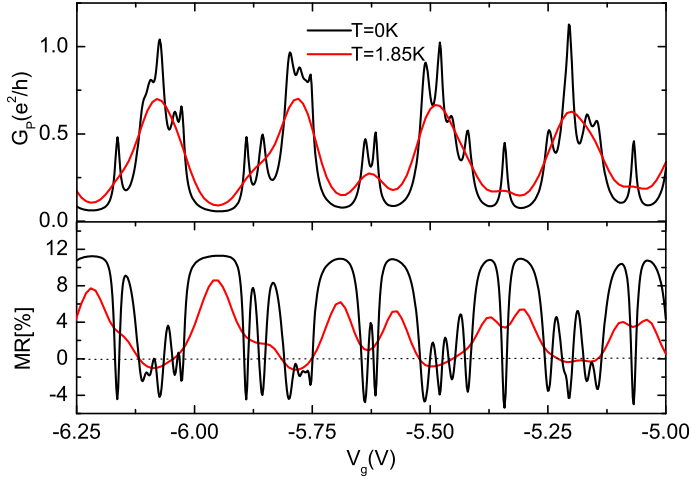}%
\caption{Conductance $G^{P}$(top panel) and magnetoresistance $MR$
(bottom panel) calculated for a MWNT\ with two ferromagnetic
contacts, for $T=0$ (black lines) and $T=1.85$ K (red lines). We
have assumed that current transport occurs through 6 independent
channels. We have used $P_{L(R)}=0.2$, $T_{L}=0.45$, $T_{R}=0.04$
and $\alpha=0.01$ for all channels. A radius of $R_{nt}=2.7nm$ and a
Fermi level of $E_{w}^{F}=1eV$ have been used for the nanotube in
order to calculate the wavevector $k_{Fw}^{n}$ associated to channel
$n\in\{1..6\}$. Beatings occur in the zero temperature signals, \
due to the combination of the different channels. Therefore, at
$T=1.85$ K, the $MR$ signal seem to oscillate with a period which is
much larger than the intrinsic
spacing between the resonant levels.}%
\label{Fig:TMRMWNTmulti}%
\end{figure}
Figure \ref{Fig:TMRMWNTmulti} shows the calculated conductance
$G^{P}$ (top panel) and the magnetoresistance $MR$ (bottom panel)
for $T=0$ (black full lines) and $T=1.85$ K (dashed red lines). We
have used contact parameters $P_{L(R)}=0.2$, $T_{L}=0.45$ and
$T_{R}=0.04$ for all the channels and a coupling $\alpha=0.01$. For
simplicity, we have assumed no SDIPS. At $T=0$, the conductance
shows peaks which correspond to the resonant states in the different
channels. Due to the larger number of channels, the intrinsic energy
spacing between these resonances is reduced. As a result of the
combination of the different conducting channels, beatings occur for
certain regions of gate voltage. Importantly, the $MR$ can become
negative, for the same reason as in the one channel case. At
$T=1.85$ K, it is not possible anymore to resolve the single level
resonances. Due to thermal averaging, the conductance and
magnetoresistance are roughly determined by the envelop of the
transmissions. Therefore, the magnetoresistance shows approximately
periodic sign changes with a period which is much larger than the
intrinsic energy spacing between the resonant states. This type of
behavior will be illustrated with measurements performed with MWNTs
in section \ref{MWNTBasel}.

\subsection{Spin dependent transport in a quantum dot.\label{EOM}}
The tunnel junctions connecting the ferromagnetic leads to the
nanotubes have often a small capacitance of the order of $10$aF. In
such a case, at low temperatures ($T<10K$), a finite charging energy
$U=e^{2}/C_{\Sigma}$ is required to add electrons on a nanotube. The
interplay of Coulomb blockade and spin dependent transport phenomena
have attracted a lot of theoretical and experimental interest
recently (see for instance
\cite{Wetzels,SETmr,Weymann,Martinek:03,Kondo,KondoHu,Cottet:04}).
We introduce below a very recent theoretical development
\cite{Cottet:06b} which allows to address spin transport in the
Coulomb blockade regime corresponding to experiment
\cite{Kontos:05}.

We assume that strong Coulomb interactions are added to the wire of
figure \ref{diagramm}, so that we have a quantum dot connected to
ferromagnetic leads. In the interacting regime, the scattering
approach used in the former section is not suitable anymore for
describing this system. One can adopt a description based on the
Anderson-like hamiltonian
\[
H=H_{dot}+H_{leads}+H_{c}%
\]
with
\begin{eqnarray}
H_{dot}  & =\sum\limits_{d,\sigma}\xi_{d\sigma}c_{d\sigma}^{\dagger}%
c_{d\sigma}+\sum\limits_{\substack{d,d^{\prime},\sigma,\sigma^{\prime}
[(d,\sigma)\neq(d^{\prime},\sigma^{\prime})]}}\frac{U}{2}n_{d\sigma}n_{d^{\prime
}\sigma^{\prime}}\label{H}\\
H_{leads}  &
=\sum\limits_{k,\sigma}\xi_{k\sigma}c_{k\sigma}^{\dagger
}c_{k\sigma}\\
H_{c}  & =\sum\limits_{d,k,\sigma}\left(  t_{d\sigma}^{k}c_{d\sigma}^{\dagger
}c_{k\sigma}+(t_{d\sigma}^{k})^{\ast}c_{k\sigma}^{\dagger}c_{d\sigma}\right)
\end{eqnarray}
Here, $\xi_{d\sigma}$ refers to the energy of the dot orbital state
$d$ for spin $\sigma$, $\xi_{k\sigma}$ to the energy of lead state
$k$ for spin $\sigma$ and $t_{d\sigma}^{k}$ is an hoping matrix
element. The index $k$ runs over the electronic states of lead $L$
and $R$. We assume that the spin $\sigma$ is preserved upon
tunneling, like in section~\ref{ballistic}. Coulomb interactions are
taken into account through the term in $U$, with
$n_{d\sigma}=c_{d\sigma}^{\dagger}c_{d\sigma}$.

Although the notion of interfacial scattering phase is less natural here than
with the scattering approach of section~\ref{ballistic}, it is possible to
take into account the effects related to the SDIPS in the present interacting
model. Indeed, by construction of hamiltonian (\ref{H}), for $U=0$, each
orbital level $\xi_{d\sigma}$ corresponds to a resonant level $E_{c}^{\sigma,j}$
of section \ref{ballistic}, with $\xi_{d\downarrow}-\xi_{d\uparrow}%
=g\mu_{B}h_{SDIPS}^{c}$. One can therefore introduce the effective
Zeeman splitting $h_{SDIPS}^{c}$ in equation (\ref{H}) as a
generalization of the SDIPS concept to the interacting case. This
can be justified physically on the following basis. In the
non-interacting case, we have considered that the ferromagnetic
exchange field leads to a spin-dependent interfacial potential,
responsible for the spin-dependent scattering. For a double barrier
system, the ferromagnetic exchange field makes the confinement
potential of electrons on the dot spin-dependent as well. This
naturally induces a spin-dependence of the orbital energies, which
is the counterpart of the spin-splitting of the resonant energies
found in section~\ref{ballistic}.

In the interacting case, the zero-bias conductance of the circuit can be
expressed as~\cite{Meir2}
\begin{eqnarray}
\frac{h}{e^{2}}\frac{G^{c}}{2}=\sum\limits_{d,\sigma}\int_{-\infty}^{+\infty
}d\omega\frac{\partial f(\hbar\omega)}{\partial\omega}\frac{\Gamma_{d\sigma
}^{L}(\hbar\omega)\Gamma_{d\sigma}^{R}(\hbar\omega)}{\Gamma_{d\sigma}%
^{L}(\hbar\omega)+\Gamma_{d\sigma}^{R}(\hbar\omega)}\Im m%
[G_{d\sigma}(\omega)]\label{cond}
\end{eqnarray}
with, for $l\in\{L,R\}$,
$\Gamma_{d\sigma}^{l}(\xi)=\sum\limits_{k}2\pi\left|
t_{d\sigma}^{k}\right|  ^{2}\delta(\xi=\xi_{k\sigma})$, and
$G_{d\sigma
}(\omega)=\int_{-\infty}^{+\infty}G_{d\sigma}(t)e^{i\omega t}dt$
with $G_{d\sigma}(t)=-i\theta(t)\left\langle \left\{
c_{d\sigma}(t),c_{d\sigma }^{\dagger}(0)\right\}  \right\rangle $.
For comparison with the experimental data of reference
\cite{Kontos:05}, one has to study current transport in the limit in
which the width of conductance peaks displayed by the circuit is not
limited only by
temperature but also by the tunnel rates ($k_{B}T<\Gamma_{d\sigma}^{L}%
+\Gamma_{d\sigma}^{R}$ ). This requires to go beyond the sequential
tunneling description (see for instance \cite{Cottet:04}), i.e. to
take also into account high-order quantum tunneling processes. For
temperatures larger than the Kondo temperature of the circuit
($T>T_{K}$), this can be done by calculating the Green's functions
$G_{d\sigma}(\omega)$ with the Equation of Motion Technique (E.O.M.)
introduced for quantum dot systems by Meir {\it et al.} \cite{Meir}.

\paragraph{Single orbital case\label{SingleEOM}}

For simplicity, we first consider a one-orbital quantum dot. Then,
the E.O.M technique leads to \cite{Meir}
\begin{equation}
\eqalign{\fl G_{d\sigma}(\omega)= \cr\frac{\hbar(1-\left\langle
n_{d\overline{\sigma}}\right\rangle )}{\hbar
\omega-\xi_{d\sigma}-\Sigma_{d\sigma}^{0}+\frac{U\Sigma_{d\overline{\sigma}%
}^{1}}{\hbar\omega-\xi_{d\sigma}-U-\Sigma_{d\sigma}^{0}-\Sigma_{d\overline
{\sigma}}^{3}}}+\frac{\hbar\left\langle
n_{d\overline{\sigma}}\right\rangle
}{\hbar\omega-\xi_{d\sigma}-U-\Sigma_{d\sigma}^{0}-\frac{U\Sigma
_{d\overline{\sigma}}^{2}}{\hbar\omega-\xi_{d\sigma}-\Sigma_{\sigma}%
^{0}-\Sigma_{d\overline{\sigma}}^{3}}}}\label{greensfunction}%
\end{equation}
where $\left\langle n_{d\sigma}\right\rangle
=-\int_{-\infty}^{+\infty}d\omega f(\hbar\omega)\Im
m[G_{\sigma}^{d}(\omega)]/\pi$ is the average occupation of orbital
$d$ by electrons with spin $\sigma$. Assuming that the coupling to
the leads is energy independent (broad band approximation), one has
$\Sigma_{d\sigma}^{0}=-i(\Gamma_{d\sigma}^{L}+\Gamma_{d\sigma}^{R})/2$,
$\Sigma_{d\overline{\sigma}}^{3}=-i(\Gamma_{d\overline{\sigma}}^{L}+\Gamma_{d\overline{\sigma}}
^{R})$ and, for $i\in\{1,2\}$,
\[
\fl\Sigma_{d\overline{\sigma}}^{i}=\sum\limits_{k}\frac{\mu_{i}(\xi
_{k\overline{\sigma}})\left|  t_{k\overline{\sigma}}^{d}\right|  ^{2}}%
{\hbar\omega-\xi_{d\sigma}+\xi_{d\overline{\sigma}}-\xi_{k\overline{\sigma}%
}+i0^{+}}+\sum_{k}\frac{\mu_{i}(\xi_{k\overline{\sigma}})\left|
t_{k\overline{\sigma}}^{d}\right|  ^{2}}{\hbar\omega-\xi_{d\sigma}%
-\xi_{d\overline{\sigma}}-nU+\xi_{k\overline{\sigma}}+i0^{+}}\,.
\]
with $\mu_{1}(\xi)=f(\xi)$ and $\mu_{2}(\xi)=1-f(\xi)$. The term
$\Sigma_{d\sigma}^{0}$, which is due to the tunneling of electrons
with spin $\sigma$, already occurred in the non-interacting case.
Indeed, for $U=0$ and $T_{l}^{\sigma}\ll1$, the conductance given by
the above equations can be perfectly mapped onto the non-interacting
conductance found in section \ref{ballistic}, using
$E_{d\sigma}^{c}=\xi_{d\sigma}$ and $T_{l}^{\sigma}=\pi
N_{Fw}\Gamma_{d\sigma}^{l}=2\pi N_{Fw}\left|
\Sigma_{d\sigma}^{0}\right|  $. In the interacting case, $G_{d\sigma}%
(\omega)$ also involves $\Sigma_{d\sigma,d^{\prime}\sigma^{\prime}}^{i,n}$
terms related to the tunneling of electrons with spin $\overline{\sigma}$.
Note that $G_{d\sigma}$, $\xi_{d\sigma}$\ and $\Gamma_{d\sigma}^{L(R)}%
$\ depend on the configuration $c\in\{P,AP\}$\ considered but for simplicity
we have omitted the index $c$\ in those quantities.

\begin{figure}[ptb]
\centering\includegraphics[width=0.55\linewidth]{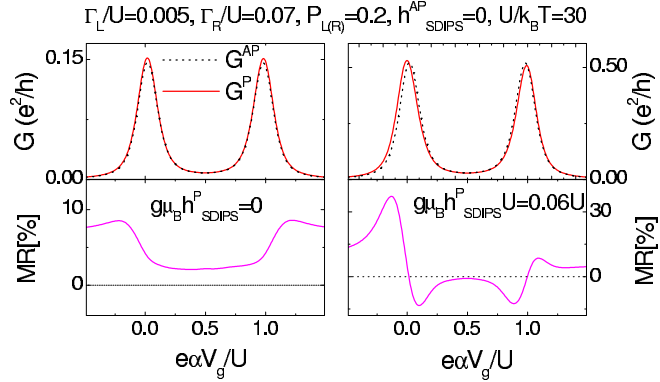}\caption{Top
panels: Conductance $G^{P}$ in the parallel configuration (red full
lines) and conductance $G^{AP}$ in the antiparallel configuration
(black dotted lines) as a function of the gate voltage $V_{g}$, for
the circuit shown in figure \ref{diagramm}, with W a 1-orbital
quantum dot. We have used $\Gamma_{L}=0.005U$, $\Gamma _{R}=0.07U$,
$P_{L(R)}=0.2$, $U/k_{B}T=30$ and $h_{SDIPS}^{AP}=0$. Bottom panels:
Magnetoresistance $MR$ (pink curves) corresponding to the above
conductance plots. The results are shown for
$g\mu_{B}h_{SDIPS}^{P}=0$ (left
panels) and $g\mu_{B}h_{SDIPS}^{P}=0.06U$ (right panels). }%
\label{EOMmono}%
\end{figure}

Figure \ref{EOMmono} shows the conductance $G^{c}$ in configuration
$c\in\{P,AP\}$ (top panels) and the magnetoresistance $MR$ (bottom
panels) calculated for different values of $h_{SDIPS}^{c}$, using
$\Gamma_{d\sigma}^{l}=\Gamma_{l}(1+\eta_{l}\sigma P_{l})$ for
$l\in\{L,R\}$. The conductance peak corresponding to level $d$ is
split by $U$ due to Coulomb interactions. For $h_{SDIPS}^{c}=0$
(left panels), we already note that although the two conductance
peaks displayed by $G^{P}(V_{g})$ are very similar, the $MR$
variations corresponding to these two peaks have different shapes
(see also \cite{Weymann}). More precisely, for the low values of
polarization considered here, $MR(V_{g})$ is approximately mirror
symmetric from one conductance peak to the other. This is in
contrast with the single channel non-interacting case in which the
$MR(V_{g})$ was identical for all conductance peaks. A finite
effective field $h_{SDIPS}^{c}$ produces a shift of the conductance
peaks from the $P$ to the $AP$ configurations. For instance, in
figure \ref{EOMmono}, top right panel, the left [right] conductance
peak is shifted to the right [left] from $P$ to $AP$ because it
comes in majority from the transport of up [down] spins in the $P$
case. As a consequence, in figure \ref{EOMmono}, bottom right panel,
the amplitude of $MR$ is enhanced and it becomes negative for
certain values of $V_{g}$. We note that using a finite SDIPS does
not help to obtain similar $MR$ variations for the two peaks of
$G^{P}(V_{g})$ because the SDIPS shifts these two peaks in opposite
directions.

Before concluding this section, we point out that, in principle, $%
h_{SDIPS}^{c}$ is not the only term which can lead to a
spin-splitting of
the dot energy levels. Indeed, the terms proportional to $\Sigma _{d%
\overline{\sigma }}^{1}$ and $\Sigma _{d\overline{\sigma }}^{2}$ in
equation (\ref{greensfunction}) can also renormalize these levels,
due to their real part. In the case of ferromagnetic contacts, this
renormalization is different for the two spin directions. This
allows to define another type of effective field, $h_{U}^{c}$, which
is intrinsically taken into account in the treatment shown here. The
effects of $h_{U}^{c}$ have been studied in detail by
\cite{Martinek:03} for a quantum dot with non-collinearly polarized
ferromagnetic leads in the sequential tunneling regime (see note
[34] of reference \cite{Cottet:06b}), and by \cite{KondoHu} for a
quantum dot in the Kondo regime. Similarly to $h_{SDIPS}^{c}$, the
value of $h_{U}^{c}$ depends on
the configuration of the ferromagnetic electrodes and it must vanish in the $%
AP$ configuration for symmetric junctions. Nevertheless, for the low
values of tunnel rates $\Gamma _{L(R)}$, polarizations $P_{L(R)}$
and the temperatures $T$ used here, $h_{U}^{c}$ is much weaker than
the finite $h_{SDIPS}^{c}$ assumed, and it can therefore not play
the same role as $h_{SDIPS}^{c}$.

\paragraph{Generalization to a non-degenerate multilevel system}

For simplicity, we have considered in the previous section the
one-orbital case. In practice, other orbital levels close to orbital
$d$ can modify the $MR(V_{g})$ pattern. Nevertheless, for non
degenerate energy levels with a sufficiently large intrinsic level
spacing $\Delta E$ (see \cite{Cottet:06b}), the two conductance
peaks associated to a given orbital will occur consecutively in
$G^{c}(V_{g})$. The SDIPS will shift these two peaks in the same way
as for the single orbital model. Therefore, one can still expect
changes of sign in the $MR(V_{g})$ curves, with dissimilar
$MR(V_{g})$ patterns for the two conductance peaks corresponding to
a given orbital level.

\paragraph{Effect of a twofold degeneracy of orbital levels.}

In single wall carbon nanotubes, a two-fold orbital degeneracy is
commonly observed, related to the K-K' energy band degeneracy of
graphene \cite{Sapmaz:05,KK}. To investigate some consequences of
this feature, one can consider a two degenerate orbitals model, i.e.
hamiltonian (\ref{H}) with $d\in \{K,K^{^{\prime}}\}$ and
$\xi_{K^{\prime}\sigma}=\xi_{K\sigma}$. For simplicity, we assume no
coupling between the two orbitals through higher orders dot-lead
tunnel processes. We also assume the same dot-lead coupling and
interfacial parameters for both orbitals. In the non-interacting
case, this modification leaves the $MR$ unchanged (see section
\ref{ballistic}). In the interacting limit, an orbital degeneracy
has more complicated effects on the $MR$. This was studied with the
E.O.M. technique in reference \cite{Cottet:06b}. We refer the
readers to this reference for details of the calculation and present
here the main results of this approach.

\begin{figure}[ptb]
\centering\includegraphics[width=0.55\linewidth]{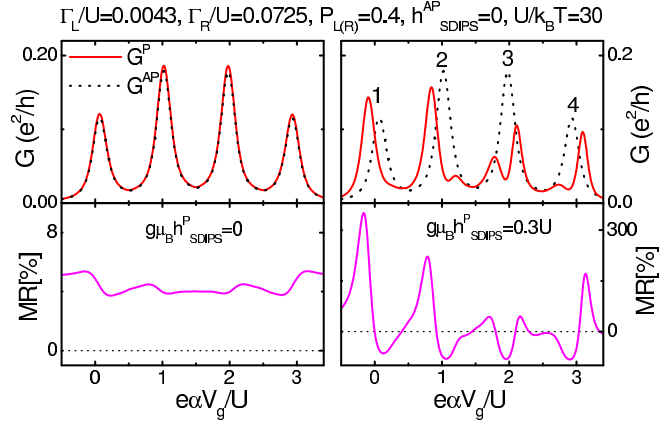}\caption{Top
panels : Conductance $G^{P}$ in the parallel configuration (red full
lines) and conductance $G^{AP}$ in the antiparallel configuration
(black dotted lines), for the circuit of figure \ref{diagramm}, with
W a two-orbitals quantum dot. We have used identical tunnel rates to
the two orbitals, i.e. $\Gamma_{L}=0.0043U$, $\Gamma_{R}=0.0725U$,
and $P_{L(R)}=0.4$. We have also used $U/k_{B}T=30$ and
$h_{SDIPS}^{AP}=0$. Bottom panels: Magnetoresistance $MR$ (pink full
lines) corresponding to the conductance plots. The results are shown
for $g\mu _{B}h_{SDIPS}^{P}=0$ (left panels) and
$g\mu_{B}h_{SDIPS}^{P}=0.3U$ (right panels). The twofold orbital
degeneracy allows to restore locally an approximate regularity of
the $MR(V_{g})$ pattern. {This behavior will be compared with
experimental results in section
\ref{SWNTBasel}.}}%
\label{EOMmulti}%
\end{figure}

Figure~\ref{EOMmulti} shows the conductance (top panels) and $MR$
curves (bottom panels) in the two orbital case, calculated for
different values of $h_{SDIPS}^{c}$. In most cases, the curves
$G^{c}(V_{g})$ show 4 resonances, the first two associated with a
single occupation of $K$ and $K^{\prime}$, and the other two to
double occupation. For $h_{SDIPS}^{P}=h_{SDIPS}^{AP}=0$ and the
parameters used here, the $MR$ remains positive for any value of
$V_{g}$ (left panels). Like in the 1-orbital case, a finite
$h_{SDIPS}^{c}$ makes easier negative $MR$ effects and allows a
stronger tunability of the $MR$ effect with $V_{g}$ (right panels).
Importantly, the effect of $h_{SDIPS}^{c}$ again depends on the
occupation of the dot and the $MR(V_{g})$ pattern is thus not
similar for the four conductance peaks. Nevertheless, in figure
\ref{EOMmulti}, top right panel, the first two conductance peaks of
$G^{P}$ (peaks 1 and 2) are both shifted to the left by
$h_{SDIPS}^{P}$ because they are both due in majority to up spins.
This allows to get a $MR$ pattern approximately similar for these
two peaks, i.e. a transition from positive to negative values of
$MR$ (bottom right panel). On the contrary, peaks 3 and 4 correspond
to a transition from negative to positive values of $MR$ because the
associated conductance peaks are due in majority to down spins. The
shape of the $MR(V_{g})$ pattern associated to the transition
between peaks 3 and 4 is particular (negative/positive/negative)
because, for the values of parameters considered here, Coulomb
blockade does not entirely suppress the up spins contribution in
peak 3. Remarkably, this allows to obtain, at the left of figure
\ref{EOMmulti}, bottom right panel, three positive $MR$ maxima which
differ in amplitude but have rather similar shapes. Taking into
account a twofold orbital degeneracy thus allows to restore an
approximate local regularity of the $MR$ pattern. Note that for
clarity, we have used in figure \ref{EOMmulti}, right panels, a
large value of $h_{SDIPS}^{P}$. Nevertheless, the effect persists
for lower values of SDIPS (see figure \ref{Fig:gateTMRSWNTBasel}).
This behavior will be compared with the experimental data of
reference \cite{Kontos:05} in section \ref{SWNTBasel}.

\subsection{Other interacting regimes.}

In sections \ref{ballistic} and \ref{EOM}, we have put a special
emphasis on the non-interacting regime and on the Coulomb blockade
regime because this is relevant for interpreting the experimental
results available so far (see section \ref{bip}). However, in
principle, a carbon nanotube connected to ferromagnetic leads can
adopt other types of behaviors. For instance, it could behave as a
quantum dot in the Kondo regime (see e.g. \cite{KondoRev}). The
effect of ferromagnetic leads on this system has been studied
theoretically by various authors \cite{Kondo, KondoHu}. A first
experimental study could be realized using $C_{60}$ molecules,
revealing a spin-splitting of the Kondo resonance related to the
coupling to the ferromagnetic leads\cite{Patsupathy:05}.
Nevertheless, in carbon nanotubes, Kondo effect has only been
observed with non-magnetic leads so far \cite{NanotubeKondo}. A
carbon nanotube can also behave as a Luttinger liquid due to the
interplay between electronic interactions and the one-dimensional
nature of the nanotube \cite{NanotubeLuttinger,Bockrath:99}. In a
Luttinger liquid, electrons form collective charge and spin
excitations which propagate with different velocities. The resulting
spin-charge separation effect remains to be observed in an
unambiguously accepted way \cite{Auslaender}. This is one more
fundamental motivation for the study of spin transport in carbon
nanotubes. It has been predicted that spin-transport could provide
experimental evidences of spin-charge separation, in the Fabry-Perot
like regime \cite{Luttinger} corresponding to section
\ref{ballistic}, as well as the incoherent regime $\max
(eV_{sd},k_{B}T)\gg h v_{FW}/\ell $ (see \cite{Balents}). We will
not report on those predictions in detail here because no
experimental realization is available at this time.

\subsection{The spin injection problem.}

Until now, we have assumed that it was possible to inject spins
reliably in carbon nanotubes. More precisely, in the framework of
the theories presented in sections \ref{ballistic} and \ref{EOM}, we
have assumed that the spin polarization $P_{l}^{n}$ of the tunnel
rates was finite. One must wonder whether this is possible in
practice.

In the last decade, the realization of spin-injection from
ferromagnetic metals into semiconductors has triggered many efforts,
motivated by the proposal by Datta and Das for a spin transistor
based on the electric field control of spin-orbit coupling through
the Rashba effect\cite{Zutic:04,Datta}. One major difficulty came
from the problem of the so-called conductivity mismatch. As shown by
\cite{vanWees:01} in the diffusive limit, when a semiconductor is in
good contact with a ferromagnetic material, the spin-polarization of
the current injected into the semiconductor is strongly reduced
because the conductivity of the semiconductor is much smaller than
that of the ferromagnet. Nevertheless, it was shown that
spin-injection can be enhanced by adding tunneling contacts at the
interfaces between the ferromagnets and the non-magnetic materials
\cite{Rashba:00,FertJaffres:01}. This turns out to be valid also in
the ballistic limit (see for instance \cite{Hu:01}), and in
particular for carbon nanotubes, as illustrated in the next section.

\subsubsection{The ballistic spin injection picture.}

Tunnel barriers are commonly obtained between ferromagnetic metals
and carbon nanotubes. This makes spin injection possible as we are
going to show here. Estimating accurately the spin injection
parameter $P_{l}^{n}$ for a ferromagnet/nanotube interface is beyond
the scope of this review. We will rather adopt here a scattering
approach with a Dirac potential barrier to model the interface. As
shown by reference \cite{Hu:01}, this simplified approach is enough
to capture some essential properties of the interfaces.

\begin{figure}[pth]
\centering\includegraphics[width=0.9\linewidth]{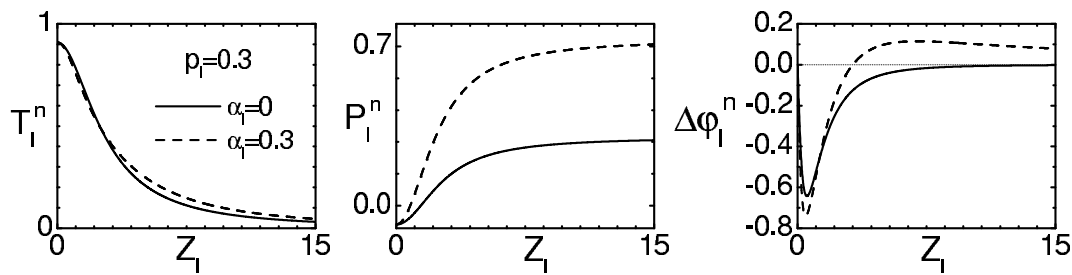}
\caption{Spin-averaged tunneling rate $T_{l}^{n}$ (left panel),
tunneling rate polarization $P_{l}^{n}$ (middle panel) and SDIPS
parameter $\Delta\varphi_{l}^{n}$ (right panel) of contact
$l\in\{L,R\}$, estimated by using a Dirac barrier model with a
spin-dependent coefficient $U_{l}^{\sigma}$, placed between a
ferromagnetic metal with Fermi wavevector $k_{l}^{\sigma}\sim 1.7 .
10^{10} m^{-1}$, and a wire with Fermi wavevector
$k_{Fw}=8.5~10^{9}\mathrm{m}^{-1}$ typical of single wall nanotubes.
We show the results as a function of the average barrier strength
$Z_{l}=m_{e}(U_{l}^{\uparrow}+U_{l}^{\downarrow})/\hbar^{2}k_{Fw}$,
for a polarization $p_{l}=0.3$ of the lead density of states and
different values of the spin asymmetry $\alpha_{l}=(U_{l}^{\downarrow}-U_{l}^{\uparrow})/(U_{l}%
^{\uparrow}+U_{l}^{\downarrow})$ of the barrier.}%
\label{Fig:spininjection}%
\end{figure}

We assume a Dirac-function potential $U_{l}^{\sigma}\delta(x-x_{l})$
for the interface $l\in\left\{L,R\right\}$ placed at $x_{l}$, and we
use a spin-dependent wavevector $k_{l}^{\sigma}$ for ferromagnetic
lead $l$. Solving the quantum mechanical problem, one finds for
junction $l$ a transmission probability:
\begin{eqnarray}
T_{l}^{n,\sigma}=\frac{4k_{Fw}k_{l}^{\sigma}}{(k_{Fw}+k_{l}^{\sigma}%
)^{2}+(2U_{l}^{\sigma}m_{e}/\hbar^{2})^{2}}\label{eq:spininjection}
\end{eqnarray}
For completeness, we also give the expression of the wire-wire
reflection phase
\[
\varphi_{l}^{n,\sigma}=\arctan\left[
\frac{2U_{l}^{\sigma}m_{e}}{\hbar
^{2}(k_{Fw}-k_{l}^{\sigma})}\right]
+\arctan\left[\frac{2U_{l}^{\sigma}m_{e}}{\hbar
^{2}(k_{Fw}+k_{l}^{\sigma})}\right]
\]
at contact $l$. Figure \ref{Fig:spininjection} shows the
spin-averaged transmission probability
$T_{l}^{n}=(T_{l}^{n,\uparrow}+T_{l}^{n,\downarrow})/2$, the
spin-polarization of the transmission probability
$P_{l}^{n}=(T_{l}^{n,\uparrow}-T_{l}^{n,\downarrow})/(T_{l}^{n,\uparrow}+T_{l}^{n,\downarrow})$
and the SDIPS parameter
$\Delta\varphi_{l}^{n}=\varphi_{l}^{n,\uparrow}-\varphi_{l}^{n,\downarrow}$
calculated from these equations. For the nanotube, we use
$k_{Fw}\sim 8.5 . 10^9 m^{-1}$ \cite{bockrath:01}, and for the
ferromagnetic contact, we use the typical value $k_{l}^{\sigma}\sim
1.7 . 10^{10} m^{-1}$ \cite{Hu:01} and a spin polarization
$p_{l}=0.3$ for the electronic density of states in lead $l$. We
define the average barrier strength
$Z_{l}=m_{e}(U_{l}^{\uparrow}+U_{l}^{\downarrow})/\hbar^{2}k_{Fw}$.
We first assume that $U_{l}^{\sigma}$ is spin-independent (full
curves). For a metallic contact, that is $Z_{l}=0$, $P_{l}^{n}$
remains very small. Nevertheless, the spin injection efficiency is
strongly enhanced for a high barrier strength \cite{Hu:01}. It is
also possible that the potential barrier between the nanotube and
the ferromagnet is itself spin-polarized, i.e. $\alpha_{l}=(U_{l}^{\downarrow}-U_{l}^{\uparrow})/(U_{l}%
^{\uparrow}+U_{l}^{\downarrow})\neq 0$. This can be due to the
magnetic properties of the contact material itself, when it is
evaporated directly on the nanotube, but it can also be obtained
artificially by using a magnetic insulator (see \cite{MI}) to form
the barrier. This allows to further enhance spin-injection (dashed
lines).

Before concluding this section, we point out that in the case of
coherent quantum transport, $P_{l}^{n}$ is not the only parameter
which sets the efficiency of spin-injection. Indeed,
$\Delta\varphi_{l}^{n}$ is also a crucial parameter since it
determines the localized quantum states inside the wire. Let us
consider for simplicity the non-interacting case of section
\ref{ballistic}. For the weak values of SDIPS used in figure
\ref{SDIPS}, spin injection was not improved. However, it was shown
that with a stronger SDIPS, the resonant states in $G^{P[AP]}$ are
spin-split, each sub-peak corresponding to a given spin-direction
(see reference \cite{Cottet:06}). Having a strong SDIPS can thus
allow to have a strongly spin-polarized current. Since the
SDIPS-induced spin-splitting is different in the $P$ and $AP$
configurations, this allows to further increase the $MR$. One can
see from figure \ref{Fig:spininjection} that the condition required
for this effect can be obtained with weakly transparent and
spin-dependent barriers (see dashed lines for $Z_{l}$ large), which
is compatible with having a large $P_{l}^{n}$.

\subsubsection{Experimental identification of spin-injection}

In sections \ref{ballistic} and \ref{EOM}, we have considered the
gate-tunable $MR$-effect produced by spin injection into the
nanotube. We will introduce in section \ref{others} other types of
$MR$ effects which are not due to spin-injection but to various
properties of the ferromagnetic contacts. For proving that
spin-injection is actually taking place in an experiment, one must
be able to discriminate the spin-injection $MR$ effect from contact
$MR$ effects.

For this purpose, one key idea is that the $MR$ found in sections
\ref{ballistic} and \ref{EOM} is mainly a \textit{two-terminal}
effect. If only one of the contacts $k\in \{L,R\}$ is ferromagnetic,
spin injection will still induce a hysteresis in $G$ as a function
of the external field $H$, but with only a very small change $\Delta
G$ when $H=\pm H_{ck}$. This slight change will occur because the
situations in
which $\overrightarrow{p_{k}}$ is parallel or antiparallel to $%
\overrightarrow{H}$ are not totally equivalent according to equation
(\ref{ResTOT}). For instance, using the same parameters as in figure
\ref{SDIPS} but with $P_{l}=0$ and $\Delta \varphi _{l}=0$ for one
of the two contacts ($l=L$ or $l=R$) and using the typical value
$g\mu _{B}H_{ck}L/\hbar v_{FW}=0.01$, one finds a value $\left|
\Delta G\right| /G<0.8\%$ much weaker than the maximum $MR$ found
for two ferromagnetic contacts in this figure. The situation seems
to be different for contact $MR$ effects: as we will see in section
\ref{others}, these effects should already be significant with a
single ferromagnetic contact if they are relevant with two
ferromagnetic contacts. We will describe in section \ref{others}
other more specific features which can allow to identify contact
$MR$ effects.

\section{Contacting carbon nanotubes with ferromagnetic contacts.}

In this section, we present the state of the art in contacting
carbon nanotubes with ferromagnetic materials. We emphasize the
characteristics of the contacts such as minimum room temperature two
probe resistance (or transmission) and the maximum $MR$ amplitude.
We also discuss contact effects which are not related to
spin-dependent transport inside the nanotubes but which could
sometimes be superimposed to the $MR$ effects described in the
previous section.

\subsection{Contacting carbon nanotubes with ferromagnetic leads.}

Contacting carbon nanotubes with metallic electrodes has been an
issue since the start of the study of their electronic properties.
Most of transport measurements have been carried out in a two probe
geometry. In the case of non-magnetic electrodes, the transmission
of the contacts determines the relevant regime for charge transport.
As we have seen above, connecting ferromagnetic contacts to carbon
nanotubes raises additional questions, like e.g. the efficiency of
spin injection. Therefore, the choice of the ferromagnet is not only
crucial regarding the coupling of the electrodes to the nanotube but
also regarding the reliability of spin injection.

In the case of non-magnetic contacts, few multi probe measurements
have been carried out \cite{Beryazdin:97,GaoBo:05} and provide new
insights to quantum transport in nanotubes. Similarly, one can
expect multi-probe measurements to provide useful information about
spin transport when some of the probes are ferromagnetic
\cite{Kim:02,Tombros:05}. We will however focus on the studies of
the two probe geometries since they are the most advanced carried
out so far. Table 1 provides a summary of these works.

The main feature to be observed in a carbon nanotube connected to
two ferromagnetic leads is a hysteresis of the resistance versus an
applied magnetic field swept in two opposite directions, like shown
in figure \ref{fig:SpinValve}, right. One delicate point is the
control of the switching of the magnetization. It turns out that it
is non-trivial to control accurately the domain pattern of the
ferromagnet nearby the contact with the nanotubes. Therefore, almost
no $MR$ curve look like the theoretical ones. The hysteresis curves
often show complex structures. As we will see, this problem
has been partially solved with $Ni_{1-x}Pd_{x}$ and $La_{2/3}Sr_{1/3}%
MnO_{3}(LSMO)$ contacts.

\subsubsection{Co contacts.}

In the pioneering work by K. Tsukagoshi and co-workers, Co contacts
were used to study spin-dependent transport in MWNTs
\cite{Tsukagoshi:99}. The two-terminal resistance of the devices at
room temperature was ranging from $8k\Omega$ to $250k\Omega$
\cite{Tsukagoshi:00}. At $4.2K$, the authors observed a $MR$ which
they attributed to spin transport along the relatively short MWNT
length ($300nm$). The magnetic field was applied in-plane. The
electrodes were both in Co and had the same geometry. Given that
there was \textit{a priori} no reason for having two different
coercive field in the two electrodes, the AP configuration was
difficult to identify. The amplitude of the hysteresis was at most
of about $+9\%$. The observed reduced value was attributed to spin
relaxation in the nanotube, yielding a spin relaxation length of
$l_{s}=260nm$. The method was extended later to SWNTs devices
\cite{Tsukagoshi:00}. Few experiments with multi-probes of Co were
carried for SWNTs \cite{Kim:02,Tombros:05}. In that case, the
two-probe resistance was ranging from about $12k\Omega$ to $M\Omega^{\prime}%
s$. The two-terminal $MR$ reported was ranging from $2\%$ to $6\%$.
In reference \cite{Tombros:05}, shape anisotropy was used to control
selectively the switching of the different Co electrodes (typical
width about 100nm). Experiments with Co electrodes were also carried
out by Zhao and co-workers \cite{Zhao:02} but a negative $MR$ of
$-36\%$ was observed.

It is important to point out that the above experiments have been
realized without a gate voltage supply $V_{g}$. In the absence of
any doping, this would imply that the carbon nanotubes were operated
at their charge-neutral point. However, it has been found that
carbon nanotubes are extremely sensitive to their chemical
environment. The chemical potential $E_{Fw}$ of a nanotube can be
strongly modified by surface adsorbates like water molecules, gas
molecules or ions \cite{Kruger:01,adsorbatesO2,Kim}. In sections
\ref{ballistic} and \ref{EOM}, we have shown that the sign and
amplitude of $MR$ strongly depends on $E_{Fw}$. This implies that
the spin signal will depend on details of the nanotube environment.
One can even expect that $MR$ differs significantly if one measures
the same sample in different cool-downs, like observed in another
experiment described in next section \cite{Jensen:05}. Therefore,
although the different signs and amplitudes of $MR$ found in the
Co/nanotube experiments are compatible with the theoretical
expectations, a further experimental investigation of the $MR$
effect requires to use a gate electrode in order to study the
dependence of $MR$ with $V_{g}$, which is much more significant than
the value of $MR$ without a gate voltage supply.

\begin{table}[ptb]%
\begin{tabular}{@{}*{8}{c}}
\br Material&SWNT&MWNT&Max $\mid
MR\mid$&MR&Gate&F-NT-N& References\\
&($k\Omega$)&($k\Omega$)&(\%)& sign &control& $\Delta G/G$ (\%)& \\
\mr Co & 15 & 8 & 36 &+ and -&no& N
&\cite{Kim:02,Tombros:05,Tsukagoshi:99,Tsukagoshi:00,Zhao:02} \cr Fe
& 80 & N & 100 &+ and -&yes& $\sim 0$ & \cite{Jensen:05,Jensen:PhD}
\cr Ni & N & N & 15 &+ and -&yes& N & \cite{Alphenaar:05} \cr NiPd &
11 & 5.6 & 17 &+ and -&yes& $<1.4$ &
\cite{Kontos:05,Morpurgo:06,Kontos:04}\cr (Ga,Mn)As &N&N&150&+ and
-&yes& $<10$ & \cite{Jensen:05,Jensen:PhD}\cr LSMO&N& 1000 & 37
&+&no&N& \cite{Hueso:06}\cr \br
\end{tabular}
\caption{Summary of the various contacting materials used so far and
their contacting properties. Columns 2 and 3 display the minimum
two-probe resistance measured at room temperature for SWNTs and
MWNTs contacted with the material indicated in column 1. Column 4
displays the maximum $MR$ amplitude measured at low temperatures.
Column 5 reports the $MR$ signs observed. Column 6 indicates whether
a gate control of the $MR$ was achieved. Column 7 indicates the
magnetic signal $\Delta G/G$ measured for nanotubes contacted with
one ferromagnetic lead and one non-magnetic lead. Column 8 indicates
the
corresponding references (N=Not reported).}%
\label{tab:contact}%
\end{table}

\subsubsection{Fe contacts.}

Fe is another possible choice for making ferromagnetic contacts on
nanotubes. There is only one study using Fe on SWNTs carried out by
A. Jensen and co-workers \cite{Jensen:05,Jensen:PhD}. In that case,
the two terminal resistances reported at room temperature vary from
$80k\Omega$ to $1M\Omega$. In this study, CVD grown SWNTs were used.
The first contacts were made on the top of catalyst squares and had
a typical size of $6\mu m\times8\mu m$. The second contact design
was two Fe electrodes with different aspect ratios, typically $10\mu
m\times300nm$ and $10\mu m\times200nm$ in order to control the
switching via shape anisotropy. However, both these contact
geometries gave similar magnetization switching for a field applied
in plane. The samples were coupled to an electrostatic gate. The
sign of the $MR$ could be changed from positive to negative with the
gate voltage. The observed $MR$ was ranging from $-50\%$ up to
$100\%$. Due to the absence of a detailed study of $MR$ versus
$V_{g}$, a clear conclusion cannot be drawn from this work.

\begin{figure}[pth]
\centering
\caption{Left : SEM micrograph of a typical F-nanotube-F
sample of Sahoo {\it et al.} \cite{Kontos:05}. NiPd contacts are
used to inject and detect spins electrically in a MWNT with a
contact separation of about 400nm. The external magnetic field is
applied in plane, either perpendicular or parallel to the axis of
the elongated NiPd strips. Right : Statistics for the contacting
scheme with NiPd on MWNTs. The typical two probe
resistance at room temperature is $20k\Omega$. }%
\label{fig:FNTF}%
\end{figure}

\subsubsection{Ni contacts.}

Ni has also been used to implement ferromagnetic electrodes on SWNTs
\cite{Alphenaar:05}. The main findings with respect to the other
works is a continuous sign change as a function of gate voltage,
from $+10\%$ to $-15\%$. Although the channel length was about
$10nm$, no size quantization was observed at 4.2K.

\subsubsection{NiPd contacts.}

In principle, all kinds of ferromagnetic alloys could be tried in
order to improve the reliability of the spin injection and/or the
switching of the magnetization. Among these choices, Pd based alloys
look particularly promising. Indeed, experiments using
$Ni_{1-x}Pd_{x}$ with $x\sim0.5$ are among the most advanced studies
for spin transport in carbon nanotubes
\cite{Morpurgo:06,Kontos:04,Kontos:05}. This choice is based on the
observation that Pd alone makes reliable contacts on MWNTs as well
as SWNTs \cite{Dai:03}. Furthermore, Pd is close to the
ferromagnetic instability with a Stoner enhancement of about $10$.
Few magnetic impurities are enough to drive it in the ferromagnetic
state (the same holds for Pt which has a somewhat lower Stoner
enhancement of about $4$). Therefore, it seems possible to combine
the good contacting properties of Pd with a finite spin
polarization. Furthermore, the use of Pd as contacting metal
prevents oxide layers from forming at the ferromagnet/nanotube
interface. This might be an advantage with respect to the methods
using pure ferromagnetic metals, because most of the ferromagnetic
oxides are anti-ferromagnetic and therefore not only depolarize the
electronic current, but also modify in general the spin activity of
the interface.

\begin{figure}[pth]
\includegraphics[width=0.95\linewidth]{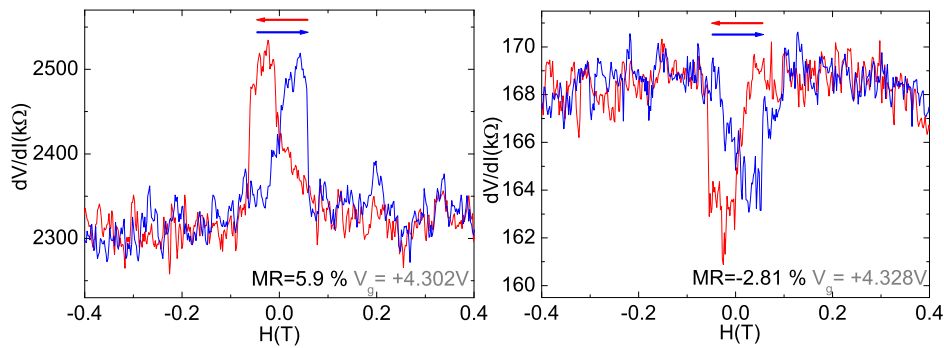}%
\caption{Example of magnetoresistance curves for the SWNT
measurements of reference \cite{Kontos:05}. Depending on the gate
voltage, both signs of the $MR$ are observed. Left : the $MR$
observed is positive ($MR=5.89\%)$, for a gate voltage
$V_{g}=4.302V$. Right : the $MR$ observed for the same device is
negative  ($MR=-2.81\%$), for a gate voltage $V_{g}=4.328V$. }%
\label{fig:HMRSWNTs}%
\end{figure}

S. Sahoo and co-workers \cite{Kontos:04,Kontos:05} were the first to
study this contacting scheme on nanotubes. The type of devices
studied is presented in figure \ref{fig:FNTF}. Two ferromagnetic
Pd$_{0.3}$Ni$_{0.7}$ strips are used to contact either a MWNT or a
SWNT. They have different shapes, typically $14$ $\mu m\times$ $0.1$
$\mu m$ and $3$ $\mu m\times$ $0.5$ $\mu m$ for the left and the
right electrode respectively. The narrower electrode has a sharp
switching around $100-250mT$. The wider one has a less pronounced
switching, as shown on figure \ref{fig:HMRSWNTs}. This suggests that
its magnetization gradually rotates upon reversing the sign of the
external magnetic field. It is worth noting that H.T. Man et al.
\cite{Morpurgo:06} as well as S. Sahoo et al. have found that the
magnetic anisotropy of the NiPd strips is in plane, perpendicular to
their long axis. This is in contradiction with the expected shape
defined anisotropy and might be related the complexity of the domain
structure of the Pd based ferromagnetic alloys. The two probe
resistance at room temperature of devices with MWNTs studied by
Sahoo et al. \cite{Kontos:04} is summarized on the right panel of
figure \ref{fig:FNTF}. As shown by this figure, the distribution of
resistances is rather peaked at the typical value of $20k\Omega$,
which shows the reliability of this contacting procedure. The
minimum value is $5.6k\Omega$, the best ever reported for
ferromagnetic contacts on MWNTs. For SWNTs, the transparency of the
contacts is lower in general, but transmission probabilities as high
as $0.84$ have been reported by H.T. Man et al. \cite{Morpurgo:06}.

At temperatures below $10K$, a $MR$ is commonly observed upon
sweeping an external magnetic field. Depending on the gate voltage,
the $MR$ is either positive or negative, as shown in figure
\ref{fig:HMRSWNTs}. The amplitude of the effect is about $5\%$ for
MWNTs and SWNTs in the ''Fabry-Perot'' regime. It grows to about
$10-15\%$ for SWNTs in the Coulomb blockade regime. In addition, the
sign of the $MR$ can also be controlled by $V_{sd}$ in both types of
nanotubes \cite{Kontos:05,Morpurgo:06}

\subsubsection{Other types of contacts.}

So far, we have only considered metallic ferromagnetic contacts to
carbon nanotubes. This choice is generally led by the simplicity of
the combination of thin metallic film growth with standard e-beam
lithography techniques. The two main drawbacks of these methods are
the small spin polarizations of the electrodes and the possible
conductivity mismatch between the metal and the carbon nanotube.

The latter issue may be solved if the ferromagnetic contacts are
ferromagnetic semiconductors. Such a method has been used recently
by Jensen \textit{et al.} who used (Ga,Mn)As ferromagnetic electrode
\cite{Jensen:05}. In this study, the Curie temperature of the
contacts is about $70K$. Therefore, the contacts are ferromagnetic
at temperatures relevant for quantum transport in carbon nanotubes,
but they cannot be used for applications at room temperature. The
maximum amplitude of the $MR$ observed at $300mK$ is however very
large, about $150\%$, and both signs of $MR$ are observed. In
addition, the sign and the amplitude of the $MR$ depend on $V_{g}$
and $V_{sd}$.

Hueso \textit{et al.} \cite{Hueso:06} have used LSMO to electrically
inject and detect spins in MWNTs. These materials have a bulk spin
polarization of about $100\%$. A $MR$ of $37\%$ is observed at $5K$,
and the spin signal persists up to $100$K. Nevertheless, this scheme
seems to produce samples with a high two-probe resistance of about
$1M\Omega$ at $300$K.

\subsection{The effect of contacts not related to spin
injection.\label{others}}

As we have seen in section \ref{coherent}, spin-polarized transport
induces a $MR$ effect for ferromagnetically contacted nanotubes.
However, a spin valve-like behavior does not automatically imply
that spin injection is actually taking place in the nanotube. This
just means that the resistance depends on the relative directions of
the magnetizations of the two ferromagnets. Although this can be
enough for spintronic devices \cite{spinFETamr:06}, from a
fundamental point of view, it is essential to separate spin
injection related phenomena from the others. In this section, we
introduce $MR$ effects not directly related to the existence of a
spin-polarized transport inside the nanotube.

\subsubsection{Stray field effects.\label{stray}}

Ferromagnetic electrodes not only induce a spin dependent scattering
at their interface but also generate an external stray field which
can be under certain circumstances of the order of a few $100mT$
\cite{Enslinn:06}. Therefore, two ferromagnetic microstrips like
shown in figure \ref{fig:FNTF} can in principle generate a local
magnetic field $H_{loc}$ which will switch hysteretically as the
magnetizations switch. Since low dimensional conductors are very
often sensitive to external magnetic fields, it is possible for a
$MR$ to appear just because charges couple naturally to the vector
potential \ (such a sensitivity is reflected in the conductance of
nanotubes connected to non-magnetic leads). Importantly, if this
mechanism is effective with two ferromagnetic contacts, it should
also be significant if only one contact is ferromagnetic, in
contrast to the spin injection case.

\begin{figure}[pth]
\includegraphics[width=0.95\linewidth]{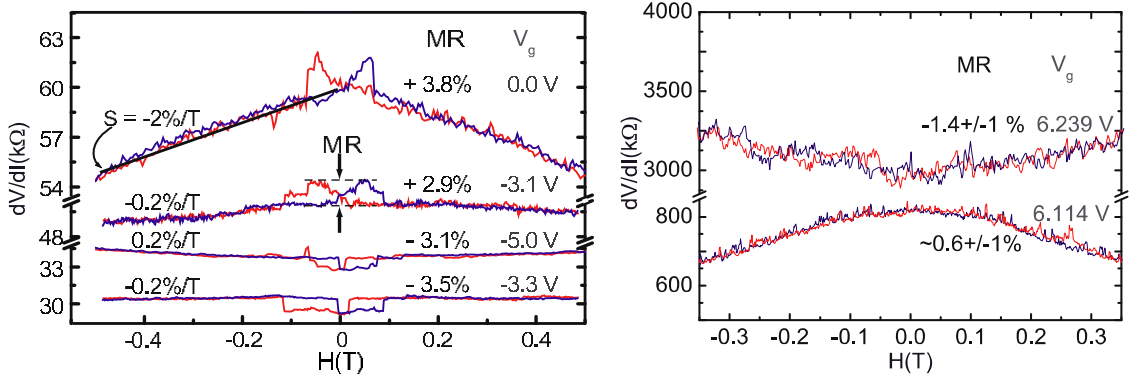}\caption{Left :
$MR$ observed by Sahoo {\it et al.} \cite{Kontos:05} for a MWNT
connected to two NiPd leads, with different values of $V_{g}$.
Depending on $V_{g}$, both signs of $MR$ and sensitivity $S$ are
observed. The amplitude and the sign of $S$ are not correlated with
the $MR$. Therefore, the potential stray fields from the
ferromagnetic electrodes cannot account for the $MR$ observed for
this device. Right : Resistance of a Pd-SWNT-PdNi device as a
function of an external magnetic field for two values of $V_{g}$.
Almost no hysteresis is observed. The maximum amplitude that can be
estimated (almost within the noise) as $\Delta G/G\sim 1\%$, more
than an order of magnitude smaller than the
observed signal with two NiPd electrodes.}%
\label{fig:contacteffects}%
\end{figure}

In practice, a MWNT with ferromagnetic contacts has indeed in
general a finite background $MR$ superimposed to the hysteretic part
of the $MR$. The field dependence of the non-hysteretic part of $MR$
can be quantified by a sensitivity $S$ in $\%/T$ to the local
magnetic field. Figure \ref{fig:contacteffects} shows typical
examples of hysteretic and background $MR$ for a MWNTs with NiPd
contacts, for different applied gate voltages $V_{g}$
\cite{Kontos:05}. The sensitivity $S$ is of the order of $1\%/T$ or
less and can change sign for different $V_{g}$. From this figure,
one can calculate the local field change $\Delta H_{loc}$ required
to obtain the observed hysteretic $MR$. For $V_{g}=-3.1V$, one finds
$\Delta H_{loc}=-2.9/0.2=-14.5T$, which is negative and way beyond
what can be obtained with microstrips. Furthermore, for
$V_{g}=-3.3V$, one would need a positive $\Delta H_{loc}$, since
both $MR$ and $S$ have the same negative sign. Such a sign change of
the local magnetic field produced by two metallic ferromagnets for
different gate voltages can hardly be explained. Therefore, stray
field effects are not dominant in the $MR$ signal for this type of
F-MWNT-F device. In addition, as one can see in figure
\ref{fig:HMRSWNTs}, $S$ is in general smaller for SWNTs
\cite{Alphenaar:05,Kontos:05,Morpurgo:06}. One can conclude that
stray field effects do not contribute substantially to the $MR$
observed in nanotubes, at least for the NiPd devices realized so
far.

\subsubsection{Magneto-Coulomb (MC) effects. \label{magneto}}

The magnetic field used to control the magnetization of the
ferromagnetic contacts can also induce a magnetoresistance effect,
independently of any spin-polarized transport process. This
so-called magneto-Coulomb (MC) effect \cite{MC} simply requires that
the conductance of the nanotube depends on its gate voltage $V_{g}$.
The magnetic field shifts the energies of spins $\uparrow$
($\downarrow)$ inside contact $l$ by the Zeeman energy $\pm
g\mu_{B}H/2$. Since the densities of states
$N_{l,\sigma}=N_{l}(1+\sigma\eta_{l}p_{l})$ in contacts
$l\in\{L,R\}$ are spin dependent ($p_{l}\neq0$), this Zeeman shift
must be compensated by a slight change
$\Delta\mu_{l}=-p_{l}g\mu_{B}H/2$ in the Fermi energies of the
contacts. This modifies the electrostatic profile of
the circuit, leading to a conductance $\widetilde{G}(V_{g})=G(V_{g}%
-C_{L}\Delta\mu_{L}/e-C_{R}\Delta\mu_{R}/e)$. In this framework, a
hysteretic conductance pattern can be obtained while sweeping the
magnetic field to reverse the magnetization of the contacts. The
amplitude of the magnetoresistance signal induced by the MC effect
is \cite{VanMolen}
\begin{eqnarray}\label{MCE}
MR=-\frac{1}{G}\frac{dG}{dV_{g}}\frac{g\mu_{B}\left(  p_{L}C_{L}H_{cL}%
+p_{R}C_{R}H_{cR}\right)  }{eC_{g}}%
\end{eqnarray}
The expression of $G$ to insert in (\ref{MCE}) depends on the
different energy scales involved in the problem. For a device
showing conductance peaks, the MC
effect should induce a magnetoresistance effect with a sign oscillating with $V_{g}%
$, since it is proportional to the derivative of $G(V_{g})$.
Importantly, equation (\ref{MCE}) shows that the MC-induced
magnetoresistance effect should occur even in the presence of one
single ferromagnetic contact. At last, from the expression of
$\widetilde{G}(V_{g})$, the MC effect produces a strong background
variation in $G(H)$ on top of the discontinuities at $H=H_{cL(R)}$
(see \cite{VanMolen}). These features could be useful for an
experimental identification of this effect.

\subsubsection{Comparison between single and double ferromagnetic contacts schemes.}

Both the stray field effect mechanism as well as the MC effect
should already be significant for devices with a single
ferromagnetic contact if these effects are relevant with two
ferromagnetic contacts. Therefore, it is useful to fabricate such
devices and measure the $\Delta G/G$. Such experiments have been
carried out by Jensen {\it et al.} with Fe-NT-Au contacts and no
$MR$ has ever been found \cite{Jensen:PhD}. With (Ga,Mn)As contacts,
Jensen {\it et al.} have reported a finite $MR$ of about $10\%$ for
single ferromagnetic contacts, while the maximum amplitude for
double ferromagnetic contacts is about $150\%$. Figure
\ref{fig:contacteffects}, right shows the $\Delta G/G$ measurement
performed by \cite{Kontos:05} for a NiPd-SWNT-Pd device, for two
different values of gate voltages, one in the Coulomb valley, and
the other close to a resonance. The upper bound for $\Delta G/G$ is
$1.4\%$ in amplitude which is one order of magnitude lower than the
maximum $\Delta G/G$ observed with two ferromagnetic contacts, as
can be seen in figure \ref{Fig:gateTMRSWNTBasel}. Therefore, all the
studies carried out so far point to the fact that contact effect are
generally not dominant.

\section{Electric field control of spin transport.}\label{bip}

In this section, we present the most advanced experimental results
which have been reported so far about the electric field control of
spin-dependent transport in carbon nanotubes. All these experiments
have been realized with NiPd contacts. For their interpretation, we
focus on the Fabry-Perot and quantum dot regimes, which have been
introduced theoretically in section \ref{coherent}.

\subsection{Spectroscopy of carbon nanotubes with ferromagnetic contacts.}

The spectroscopy of a carbon nanotube contacted to ferromagnetic
leads can be realized by measuring its conductance as a function of
the gate voltage $V_{g}$ and the source-drain voltage $V_{sd}$. This
step is essential to determine the different characteristic energies
which set the behavior of the nanotube and understand the physics
leading to the $MR$ effect. In the Fabry-Perot regime, the
spectroscopy reveals the intrinsic energy spacing $\Delta
E=hv_{Fw}/2\ell$ of the quasi bound-states of the nanotube, where
$\ell$ is the effective nanotube length on which transport is
actually taking place (see figure \ref{CondPlot.eps}, left panel).
This length is generally defined by the inner spacing between the
two metallic electrodes for SWNTs (see for instance
\cite{Sapmaz:05}) but can also be related to the full tube length
for MWNTs \cite{Buitelaar:02}. In case of a quantum dot behavior,
the spectroscopy also reveals the charging energy
$U=e^{2}/C_{\Sigma}$ of the nanotube device (see figure
\ref{Fig:gateTMRSWNTBasel}, left panel). In the latter case, other
energy scales can be revealed in the fine structure of spectroscopy
in metallic SWNTs \cite{Sapmaz:05}, but we will omit them since they
have not been identified in the F-SWNT-F experiments so far.

\subsection{Gate modulations of the magnetoresistance}

We have shown in section \ref{coherent} that the $MR$ of a quantum
wire with ferromagnetic contacts can strongly depend on the gate
voltage $V_{g}$, in the Fabry-Perot regime as well as in the quantum
dot regime. We show below how these phenomena can be revealed in
ferromagnetically contacted MWNTs and SWNTs.

\subsubsection{SWNTs in the Fabry-Perot regime\label{SWNTDelft}}

Man {\it et al.} \cite{Morpurgo:06} have measured the $MR$ for SWNTs
with transparent NiPd contacts. In agreement with previous studies
with non-ferromagnetic contacts, the characteristic pattern of an
electronic interferometer is observed, as shown on figure
\ref{CondPlot.eps} left panel. The pattern reveals an intrinsic
level spacing $\Delta E\sim7meV$, which corresponds to
zero-dimensional states delocalized over $300nm$, in agreement with
the lithographically defined SWNT length. Figure \ref{CondPlot.eps},
right panel displays simultaneous measurements of the $MR$ and the
linear conductance $G^P$. The $MR$ oscillates from $0\%$ to $4\%$ on
the same gate scale as the linear conductance, e.g. $\Delta V_g
\approx 0.5V$. Therefore, as anticipated from section
\ref{ballistic}, the $MR$ can be gate controlled in SWNTs through
quantum interferences.

\begin{figure}[pth]
\centering%
\caption{Experimental results obtained by reference
\cite{Morpurgo:06} with a SWNT connected to two PdNi contacts. The
left panel shows a colorscale plot of the non-linear conductance
$G^{P}$ of the nanotube as a function of the source-drain voltage
$V_{sd}$ and the gate voltage $V_{g}$. The spacing between the
resonant lines of conductance indicate an intrinsic energy spacing
of the levels by $\Delta E\approx 7meV$. The right panel shows a
comparison between the $G^{P}(V_{g})$ and $MR(V_{g})$ data measured
at $T=4.2K$ (symbols) and the non-interacting scattering theory of
section \ref{1channel}, assuming two uncoupled channels with
$T_{L}=0.84$, $T_{R}=0.26$, $P_{L(R)}=0.1$ and no SDIPS.}%
\label{CondPlot.eps}%
\end{figure}

In order to rule out possible contact effects, one can compare the
experimental $MR$ measured by \cite{Morpurgo:06} with the $MR$
expected from the magneto-Coulomb effect. In principle, the
so-called magneto-Coulomb (MC) effect can occur even in a
non-interacting resonant wire since the conductance of the wire
depends on its gate voltage $V_{g}$. The amplitude expected for the
MC-induced magnetoresistance, using equation (\ref{MCE}) with
$C_{g}/C_{\Sigma}=0.014$, $G(dG/dV_{g})^{-1}=0.125V $,
$H_{cL(R)}<300$ mT, $p_{L(R)}=0.1$, is $\left|  MR\right| <0.2\%$.
This value is much weaker than the measured $MR$. Also, the
experimental $MR(V_{g})$ signal is clearly not proportional to the
logarithmic derivative of $G(V_{g})$, in contrast to the $MR$
expected from equation (\ref{MCE}). It is thus not possible to
attribute the $MR(V_{g})$ observed to the magneto-coulomb effect
introduced in section \ref{magneto}.

\begin{figure}[pth]
\centering%
\caption{Experimental results obtained by reference \cite{Kontos:05}
with a MWNT connected to two PdNi contacts. Panel (a) shows the $MR$
data measured at $T=1.85$ K. The $MR$ oscillates with a period
$\Delta V_{g}^{TMR}\sim0.4-0.75$ V. Panel (b) shows a colorscale
plot of the non-linear conductance $G^{P}$ of the nanotube as a
function of the source-drain voltage $V_{sd}$ and the gate voltage
$V_{g}$, for $T=300$ mK. This plot allows to resolve the single
electron states, which correspond to a gate voltage scale $\Delta
V_{g}^{e}\sim25$ mV, and indicates Coulomb blockade effects. Panel
(c) shows the conductance $G^{P}$ of the device measured at $T=300$
mK on a $V_{g}$ range much larger than Panel (b). The conductance
peaks show beatings with a period $\Delta V_{g}^{beat}\sim0.4$ V
comparable to $\Delta V_{g}^{TMR}$.
}%
\label{Fig:gateTMRMWNTBasel}%
\end{figure}

Resonant effects account much better for the observed $MR$. Indeed,
Man \textit{et al}.~\cite{Morpurgo:06} have interpreted their data
with the model presented in section \ref{ballistic} [equations
(\ref{Gballist}) and (\ref{ResTOT})]. They have assumed two
uncoupled identical channels in order to take into account the
two-fold degeneracy commonly observed in SWNTs, with $T_{L}=0.84$,
$T_{R}=0.26$, $P_{L}=P_{R}=0.1$ and no SDIPS. In view of the strong
value of $T_{L}+T_{R}$ and of the low values of $P_{L(R)}$, the
effects of the SDIPS on the $\mathrm{MR}(V_{g})$ curves are indeed
probably too weak to be resolved in the actual experiment.
Nevertheless, it is interesting to note that the $MR(V_{g})$ pattern
of figure \ref{CondPlot.eps}, right panel,  shows a slightly
asymmetric behavior for $V_{g}<7.7$ \textrm{V}, similarly to the
curve shown in the
bottom right panel of figure \ref{SDIPS}, plotted for $T_{L}=0.84$, $T_{R}=0.26$, $P_{L}%
=P_{R}=0.1$ and a finite SDIPS value $\Delta\varphi_{L(R)}=-0.035$.
The irregularities present in the variations of the $MR(V_{g})$ data
for $V_{g}>7.7$ \textrm{V} prevent from concluding reliably on the
presence of SDIPS in these data (the authors suggest that these
irregularities are due to the the misorientation of the
magnetizations in the electrodes).

\subsubsection{MWNTs.\label{MWNTBasel}}

Sahoo et al. \cite{Kontos:05} have studied the gate dependence of
the $MR$ for MWNTs with NiPd electrodes, at $T=1.85$ \textrm{K}. As
shown in figure \ref{Fig:gateTMRMWNTBasel}a, the $MR$ is observed to
oscillate relatively regularly between \mbox{$-5$\,\%} and
\mbox{$+6$\,\%} on a gate-voltage scale $\Delta V_g^{TMR}$ such that
$0.4V<\Delta V_g^{TMR}<0.75V$.

The conductance of the same sample has been studied at lower
temperatures ($T=300$ \textrm{mK}), in order to resolve the
single-electron states which could not be resolved at the
temperature at which the $MR$ was measured. A measurement of the
differential conductance $dI/dV$ as a function of source-drain
$V_{sd}$ and gate voltage $V_{g}$ at $T=300$ \textrm{mK} is shown in
figure \ref{Fig:gateTMRMWNTBasel}-b for a relatively narrow $V_{g}$
range. It displays the diamond-like pattern characteristic for
single-electron tunnelling in a quantum dot. The diamonds vary in
size with single electron addition energies ranging between $0.5$
and \mbox{$0.75$\,meV}, in agreement with previous reports on MWNT
quantum dots with non-ferromagnetic leads \cite{Buitelaar:02}. The
$MR$ gate-voltage scale $\Delta V_{g}^{TMR}$ measured at $T=1.85$
\textrm{K }is much larger than the scale $\Delta V_{g}^{e}\sim25mV$
for the addition of single electrons: it corresponds to the addition
of at least $16$ electrons rather than $1$.

In order to understand this discrepancy, one can consider the
linear conductance observed over a wider gate-voltage range, as
shown in figure \ref{Fig:gateTMRMWNTBasel}-c. The single-electron
conductance peaks are strongly modulated in amplitude, leading to a
regular beating pattern with a gate-voltage scale  $\Delta
V_{g}^{beat}\sim0.4$ \textrm{V}. This scale corresponds to the scale
$\Delta V_{g}^{TMR}$ of the $MR$ oscillations, probably because, due
to thermal averaging at $T=1.85$ \textrm{K}, the conductance is
determined by the envelope of these beatings, which affects in turn
the magnetoresistance.

Interestingly, such beatings can be found within the multi-channel
non-interacting picture introduced in section \ref{multichannel}. In
this model, at temperatures such that the single-particle resonances
are averaged out, the $MR$ is only sensitive to the average over
these resonances, yielding a $MR$ modulation that follows the
envelope function of the single-electron peaks (see figure
\ref{Fig:TMRMWNTmulti}).

\subsubsection{SWNTs in the Coulomb blockade regime\label{SWNTBasel}}

\begin{figure}[pth]
\centering\includegraphics[width=0.45\linewidth]{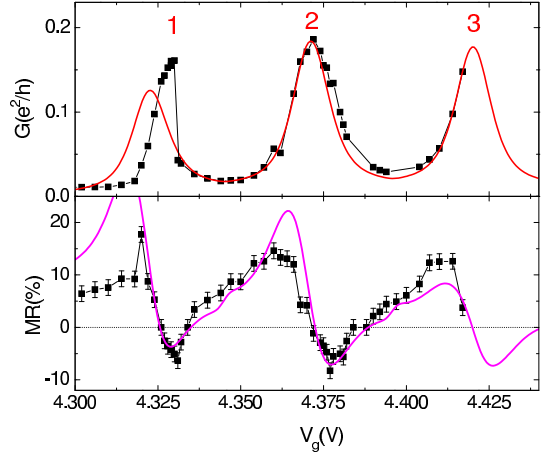}
\caption{Experimental results obtained by reference \cite{Kontos:05}
with a SWNT connected to two PdNi contacts. Left panel: colorscale
plot of the non-linear conductance $G^{P}$ of the nanotube as a
function of the source-drain voltage $V_{sd}$ and the gate voltage
$V_{g}$. This plot indicates an intrinsic energy spacing of the
levels by $\Delta E=2.5meV$ and a charging energy $U=5meV$. Right
panel: conductance $G^{P}$ and magnetoresistance $MR$ measured
simultaneously at $T=1.85K$ (black squares). As shown by reference
\cite{Cottet:06b}, these curves can be interpreted in an interacting
picture by using the E.O.M approach presented in section \ref{EOM}
for a quantum dot with two degenerate energy levels. The theoretical
curves are shown for parameters consistent with the experiment, i.e.
$U=5~\mathrm{meV}$, $U/k_{B}T=30$, $\alpha=0.0986$ and
$P_{L(R)}=0.4$. Assuming identical tunnel couplings for the two
orbitals, the values of tunnels rates $\Gamma_{L}=0.0043U$ and
$\Gamma_{R}=0.0725U$ are
imposed by the width and height of the conductance peaks. Then, $h_{SDIPS}%
^{P[AP]}$ are the only truly free fitting parameters remaining for
interpreting the $MR$ curve. The theory (colored full lines) is
plotted here for $g\mu_{B}h_{SDIPS}^{P}=0.05U$ and
$h_{SDIPS}^{AP}=0$.}%
\label{Fig:gateTMRSWNTBasel}%
\end{figure}

Sahoo et al. \cite{Kontos:05} have also studied the $MR$ for SWNTs
with NiPd contacts. Figure \ref{Fig:gateTMRSWNTBasel} left panel
displays the color plot of the non-linear conductance $dI/dV$ as a
function of $V_{g}$ and $V_{sd}$ at 1.85K for a SWNT device with
NiPd electrodes. The characteristic quantum dot behavior is
observed. One has $Ec\sim5meV$ and $\Delta E\sim2.5meV$. The latter
value corresponds to zero-dimensional states delocalized on
$\ell=600nm$, in agreement with the lithographically defined SWNT
length. In figure \ref{Fig:gateTMRSWNTBasel} right panel, the
variations of the linear conductance $G$ and the $MR$ are
simultaneously shown for two resonances. The $MR$ changes sign on
each conductance resonance. The amplitude of the $MR$ ranges from
\mbox{$-7$\,\%} to \mbox{$+17$\,\%}, which is a higher amplitude
than for the MWNT samples and SWNTs in the strongly coupled regime.
Electron-electron interactions seem to enhance the amplitude of $MR$
modulations, thereby improving the spin-FET behavior.

In this paragraph, we compare the experimental $MR$ reported in
\cite{Kontos:05} with the $MR$ expected from the magneto-Coulomb
effect. One can evaluate the amplitude of the magnetoresistance
induced by the MC effect in this experiment, using equation
(\ref{MCE}). With $C_{\Sigma}/C_{g}=10$, $G(dG/dV_{g})^{-1}=12.5 $mV
, $H_{cL(R)}<100$ mT, $p_{L(R)}=0.4$, one obtains an amplitude
$\left| MR\right| <0.4\%$, which is too weak to account for the data
of figure \ref{Fig:gateTMRSWNTBasel}. Also, the shape of the
$MR(V_{g})$ shown in this figure is clearly not proportional to the
logarithmic derivative of $G(V_{g})$, contrarily to what is expected
for the MC-induced magnetoresistance. At last, circuits with a
single ferromagnetic contact were also realized in order to check
the origin of the $MR$ effect observed. With a single ferromagnetic
contact, the $MR$-effect obtained (see figure
\ref{fig:contacteffects}) is much weaker, which rules out the MC
effect but also stray field effects produced by the ferromagnetic
leads. Therefore, one can consider the $MR$ observed with two
ferromagnetic contacts as an effect of spin-injection in a resonant
system.

In reference \cite{Kontos:05}, Sahoo et al. have used the scattering
approach introduced in section \ref{ballistic} in order to interpret
their data. The line shape of the $MR$ dips is asymmetric, similarly
to the calculated line shape for the non-interacting regime
displayed in figure \ref{SDIPS} left panel. This suggests that a
finite SDIPS can be observed in this circuit. Nevertheless, an
interacting approach which takes into account Coulomb blockade is
required in order to confirm this point. We discuss below a fully
interacting approach which allows to fit quantitatively the data, as
shown in figure \ref{Fig:gateTMRSWNTBasel}, right panel.

Reference \cite{Cottet:06b} has provided an interacting
interpretation of the data, using the E.O.M. approach presented in
section \ref{EOM}. The regularity of the $MR(V_{g})$ oscillations
displayed by the data being incompatible with a 1-orbital model, a
two-degenerate-orbital model, which takes into account the K-K'
degeneracy the orbital levels of the nanotube, has to be used. The
two-orbital model exhibits a good agreement with the experimental
data for $h_{SDIPS}^{P}=0.05U$, $h_{SDIPS}^{AP}=0$,
$\Gamma_{L}/U=0.0043$, $\Gamma_{R}/U=0.0725$, $\left|
P_{L(R)}\right|  =0.4$, $U=5~\mathrm{meV}$\textrm{,} $U/k_{B}T=30$,
and $\alpha=0.0986$. Note that the two-orbital model could not
provide a reasonable fit to the data for $h_{SDIPS}^{P[AP]}=0$.

The value of $h_{SDIPS}^{P}$ for the best fit corresponds to a
magnetic field of about $2~\mathrm{T}$, which is too strong to be
attributed to stray fields from the ferromagnetic electrodes (see
section \ref{stray}). For comparison, one can estimate
$h_{SDIPS}^{P}$ in the non-interacting theory \cite{Cottet:06},
using realistic parameters i.e leads with a Fermi energy
$10$~\textrm{eV} and a density of states polarized by $40\%$, and a
nanotube with Fermi wavevector $8.5$~$10^{9}\mathrm{m}^{-1}$, Fermi
velocity \cite{bockrath:01} $v_{Fw} =8$~$10^{5}\mathrm{m.s}^{-1}$,
length $\ell=500$~\textrm{nm} like in reference \cite{Kontos:05},
and density of states $N_{Fw}=2\ell/\pi\hbar v_{Fw}$. The interfaces
between the nanotube and the leads are furthermore modeled with
Dirac potential barriers, with a height which is spin-polarized by
$40\%$ and an average value which corresponds to (see section
\ref{SingleEOM}) $\Gamma_{L(R)}=$ $T_{L(R)}/2\pi N_{F}^{L(R)}\sim
60$~$\mathrm{\mu}$\textrm{eV }(For comparison the fitting parameters
used in
Fig.~\ref{Fig:gateTMRSWNTBasel} correspond to $\Gamma_{L}=21$~$\mathrm{\mu}%
$\textrm{eV} and $\Gamma_{R}=362$~$\mathrm{\mu}$\textrm{eV}). This gives
$h_{SDIPS}^{P}\sim1.3~\mathrm{T}$, which is consistent with the above value
used for the fit.

Note that the fitting curves shown in figure
\ref{Fig:gateTMRSWNTBasel} have been optimized in order to interpret
the data for $V_{g}>4.331~$\textrm{V}. Like many Coulomb blockade
devices, the nanotube circuit studied in this experiment suffered
from low frequency $V_{g}$-noise, which can be attributed to charge
fluctuators located in the vicinity of the device. At
$V_{g}=4.331~$\textrm{V}, a gate voltage jump occured. Therefore,
one cannot be sure that the data for $V_{g}>4.331~$\textrm{V} and
$V_{g}<4.331~$\textrm{V} correspond to the filling of consecutive
levels. Nevertheless, there is a certain probability that this is
the case since these gate voltage jumps have often an amplitude
which does not exceed $e/C_{g}$. In this case, the discrepancy
between the theory and the data could be due to the presence of
other levels which should modify the theory for peak 1. In future
experiments, it would be interesting to obtain continuous data on a
larger $V_{g}$-range, in order to check that the shape of the
$\mathrm{MR}(V_{g})$ pattern depends on the occupation of the dot (a
different shape is expected for peak 4 in the theory of Ref.
\cite{Cottet:06b}).

\subsection{Effect of source-drain bias on the magnetoresistance}

The effect of source-drain bias $V_{sd}$ on the $MR$ can also be
investigated in order to obtain a further understanding of the
system. The $MR$ at finite bias can be defined as
$(dV/dI_{AP}-dV/dI_{P})/(dV/dI_{P})$.

\begin{figure}[pth]
\centering\includegraphics[width=0.65\linewidth]{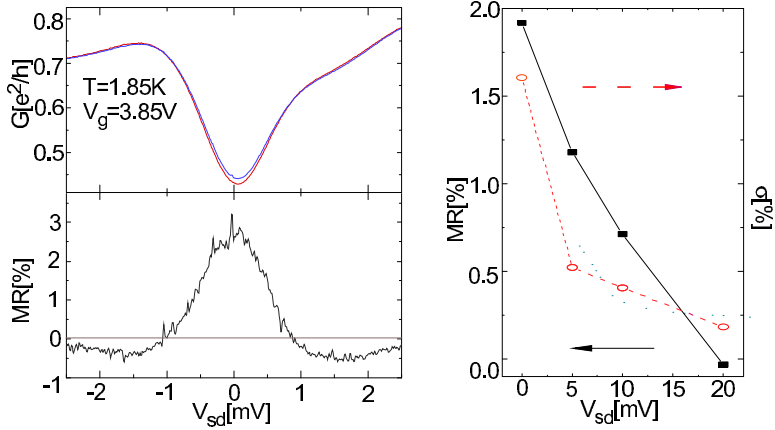}%
\caption{Effect of a finite source-drain voltage $V_{sd}$ on the
magnetoresistance of a carbon nanotube. The left panel corresponds
to a MWNT study by Sahoo {\it et al.}\cite{Sahoo:PhD}. The $MR$ is
displayed below the corresponding conductances in the $AP$ (red full
line) and the $P$ orientations (blue full line), for a finite gate
voltage of $3.85V$. The right panel corresponds to the SWNT study of
Man {\it et al.} \cite{Morpurgo:06}. $MR$ corresponds here to a gate
voltage averaged $MR$ and $\sigma$ is the corresponding standard
deviation. In both cases, the $MR$ signal vanishes when the source
drain voltage increases.}%
\label{Fig:TMRbias}%
\end{figure}

Figure \ref{Fig:TMRbias} displays two examples of $MR$ as a function
of $V_{sd}$. The left panel shows a measurement by S. Sahoo
\cite{Sahoo:PhD} of a MWNT with NiPd contacts separated by $1 \mu m$
(this is a different MWNT sample than the one discussed in section
\ref{MWNTBasel}). The $MR$, which is about $3\%$ at zero bias,
gradually decreases at finite bias and vanishes for
$V_{sd}>3mV$. It displays a sign change, symmetrically for $\mid V_{sd}%
\mid=1mV$. This energy scale corresponds to the Zero Bias Anomaly
(ZBA) observed in the conductance data shown above the $MR$. This
ZBA has been reported in MWNTS and SWNTs. In the latter case, it has
been attributed to Luttinger Liquid behavior \cite{Bockrath:99}. In
MWNTs, the ZBA has been attributed to the interplay of
electron-electron interactions and disorder \cite{Bachtold:01}.
Figure \ref{Fig:TMRbias}, right panel, shows the gate voltage
averaged $MR$ of a SWNT measured by Man {\it et al.}
\cite{Morpurgo:06}(this is the sample introduced in section
\ref{SWNTDelft}). A similar trend as for MWNTs is observed.

Within the non-interacting picture, the $MR$ should display a
similar dependence versus finite bias as versus gate voltage.
However, the general features found experimentally contradict this
simple assumption. The discrepancy between the non-interacting model
and the data of figure \ref{Fig:TMRbias} might therefore be due to
interactions. In the interacting case, Coulomb blockade can induce a
non-trivial dependence of $MR$ versus $V_{sd}$ \cite{SETmr}. For
instance, Dynamical Spin Blockade is expected to strongly affect
spin transport at finite bias in quantum dots in the sequential
tunneling regime \cite{Cottet:04}. Non trivial variations of
$MR(V_{sd})$ are also expected in the cotunneling regime
\cite{Weymann}. Furthermore, oscillations of the $MR$ signal are
predicted in the Luttinger Liquid limit \cite{Balents}. However, the
fact that the $MR$ does not seem to saturate to its classical value
(i.e. that for two tunnel junctions in series) but rather vanishes
at high bias points to possible spin relaxation processes in the
nanotube or to a bias dependence of the spin polarization in the
electrodes \cite{Morpurgo:06}.

\subsection{Spin relaxation time}

We have reported above on various experiments which indicate that a
carbon nanotube can convey a spin-polarized current. The success of
these experiments relies on the fact that electrons have a
sufficiently long spin relaxation time $\tau _{s}$ inside the
nanotube. Nevertheless, for realizing a further control of the spin
dynamics in nanotubes, a detailed study of spin relaxation processes
would be useful.

On the theoretical side, few predictions for the value of $\tau
_{s}$ in carbon nanotubes are available \cite{Alloul:05}. In thin
films made out of usual metals like Cu, one finds $\tau _{s}\sim
10ps$ at low temperatures \cite{vanWees:03} due to mechanisms
involving spin-orbit coupling and momentum scattering\cite
{Zutic:04}. In principle, one can expect a much larger $\tau _{s}$
in nanotubes, due to the very weak spin-orbit interaction expected
\cite{Egger:05,Ando:00} and the possible ballistic transport in
these systems. Electronic confinement should further suppress
conventional spin relaxation processes, as shown by
\cite{Khaetskii:00} for GaAs quantum dots. One can thus expect that
the dominant intrinsic relaxation mechanism arises from hyperfine
coupling to the nuclear spins\cite{Zutic:04}. Nevertheless, the
latter mechanism may not be so critical since $^{12}C$ does not have
a nuclear spin and $^{13}C$, which has a nuclear spin $I=1/2$, has a
low natural abundance of $1.1\%$. Very recently, Semenov \textit{et
al.} estimated $\tau _{s}\sim 1s$ due to the hyperfine interaction,
for semiconducting SWNTs at $T=4K$ \cite{Alloul:05}.

On the experimental side, two different types of methods can be used
to measure $\tau _{s}$ in nanotubes : spin injection methods and
spectral methods. Regarding spin-injection, spin must be conserved
for at least the dwell time of the electron on the nanotube in order
to produce a finite spin signal in the
conductance of the whole device. This allows to estimate a lower bound for $%
\tau _{s}$ from the experiments reported in this review. From the
measurements on SWNTs with NiPd contact in the weakly coupled regime
\cite{Kontos:05}, one finds $\tau _{s}>2\hbar /(\Gamma _{L}+\Gamma
_{R})\sim 3ps$ at $T=1.85K$. Regarding spectral methods, Conduction
Electron Spin Resonance (CESR) has been used to investigate the spin
relaxation processes in macroscopic amounts of carbon
nanotubes\cite{Petit:97,Salvetat}. So far, no consensus has emerged
from these measurements, especially concerning chemically undoped
SWNTs. However, in all these experiments, it is found that magnetic
impurities (probably catalytic particles) dominate the signal in
general for unpurified nanotubes. For vacuum-annealed SWNTs, Petit
et al. could restore a finite CESR signal and determine a relatively
long $\tau _{s}$ of $3-5ns$ at $T=300K$. Nevertheless this result
has not been reproduced in later experiments \cite{Salvetat}.
Further research on spin relaxation mechanisms in nanotubes is
highly desirable.

\section{Conclusion and perspectives}

In this review, we have shown that carbon nanotubes are promising
candidates for the realization of efficient spin-transistors.
Ferromagnetic contacts can be used to inject a spin polarized
current inside the nanotube, allowing to observe a spin-valve
behavior. A gate-tunability of the nanotube magnetoresistance has
been observed, in agreement with theoretical predictions made for
resonant tunneling systems and quantum dots.

From a technical point of view, the presently most advanced
experiments regarding the gate-control of the magnetoresistance are
not those which show the most efficient spin-injection. An
optimization of the contact properties has still to be done in this
kind of experiment, in order to obtain an accurate gate control of
the giant magnetoresistance effect. Experiments with highly
polarized ferromagnetic materials should be further developed in
order to increase the efficiency of spin injection and thus the
amplitude of the $MR$ effect. Another possibility to investigate is
using ferromagnetic insulators as tunnel barriers. The shape of the
ferromagnetic contacts should also be optimized in order to get a
better control of the switching behavior of the magnetic
polarizations. Another problem is the low temperatures required in
order to obtain discrete levels on the nanotubes. In order to
increase the operating temperature of the carbon nanotube based
spinFET, one could reduce the spacing between the ferromagnetic
electrodes down to few 10nm, as suggested very recently
\cite{Zutic:05}.

From a fundamental point of view, a more extensive study of the
dependence of the nanotube magnetoresistance on the gate voltage,
the source drain voltage and the temperature would allow to refine
the understanding of the physics involved. For instance, it would be
interesting to investigate the effects of the gate voltage on the
contacts scattering properties. It would also be interesting to
study how the Spin-Dependence of Interfacial Phase Shifts varies
with the polarization of the contacts. Nanotube spin-valves could
also be used in non-collinear configurations in order to study
spin-precession effects. Eventually, the relevant spin relaxation
mechanisms should be identified.

The studies introduced in this review open a path to the control and
the manipulation of spin in nanotubes. Besides to spintronics
applications, we believe that devices such as the ones depicted here
could also prove to be useful for quantum computing applications.

\ack A.C. and T.K. acknowledge fruitful discussions with G. Bauer,
R. Egger, A. Fert, H. Jaffr\`es, P. Se\~{n}eor C. Strunk and H.S.J
van der Zant. H.T.M. and A.F.M. gratefully acknowledge FOM and NWO
(Vernieuwingsimpuls 2000 program) for finacial support. M.-S.C. is
supported by the SRC/ERC program (R11-2000-071) and the KFR Grant
(KRF-2005-070-C00055). This work was financially supported by the
RTN Spintronics, DIENOW, by the Swiss NSF and the NCCR Nanoscience.
A.C. is financially supported by R\'egion Ile-de-France.

\end{document}